\documentstyle[stwol,epsf]{article}

\input{psfig}
\def\bzbzbar{B^0\bar B^0}
\def\Dstomunu{D_s^+\rightarrow\mu^+\nu_{\mu}}
\def\Dstotaunu{D_s^+\rightarrow\tau^+\nu_{\tau}}
\def\to{\rightarrow}
\def\qsq{q^2}

\def\btodstlnu{\bar B\to D^*\ell^-\overline\nu}
\def\bztodstlnu{\bar B^0\to D^{*+}\ell^-\overline\nu}
\def\bmtodstlnu{B^-\to D^{*0}\ell^-\overline\nu}
\def\btodststlnu{\bar B\to D^{**}\ell^-\overline\nu}
\def\btodlnu{\bar B\to D\ell^-\overline\nu}
\def\bztodlnu{\bar B^0\to D^+\ell^-\overline\nu}
\def\btopilnu{\bar B\to\pi\ell^-\overline\nu}
\def\btoclnu{b\to c\ell^-\bar\nu}
\def\btoulnu{b\to u\ell^-\bar\nu}
\def\dtokpi{D^0\to K^-\pi^+}
\def\dstophipi{D_s^+\to\phi\pi^+}
\def\slbr{B_{\rm SL}}

\def\be{\begin{equation}}
\def\ee{\end{equation}}
\def\bea{\begin{eqnarray}}
\def\eea{\end{eqnarray}}

\begin{document}

\title{PROGRESS IN UNDERSTANDING HEAVY FLAVOR DECAYS}

\author{JEFFREY D.~RICHMAN}

\address{Department of Physics, University of California, 
Santa Barbara, CA 93106}


\vskip 3mm
\twocolumn[\maketitle\abstracts{
I review new results on particles containing a charm or bottom
quark, focusing on measurements that give insight into the dynamics
of the decay process.
Leptonic and semileptonic decays are the simplest modes, and they
provide detailed tests of theoretical predictions based on
methods such as lattice QCD, heavy quark 
effective theory, and QCD sum rules. 
Although hadronic decays 
are much more complicated, the factorization 
hypothesis makes predictions that, at least for certain 
processes, are in accord with measurements. 
I also emphasize the importance of precise
measurements of branching fractions for
normalizing modes, whose uncertainties 
propagate into many other quantities.
Rare hadronic decays are now becoming accessible to
several experiments, and I discuss new results and their implications.  
Finally, I review $b$-hadron lifetime measurements, which are
steadily improving in precision and which indicate a significant 
difference between $B$-meson and $b$-baryon lifetimes.
}]

\section{Introduction}

The weak decays of hadrons containing a charm ($c$) or 
bottom ($b$) quark provide insight into a 
broad range of questions in particle physics. The main
issues are (1) decay dynamics, especially the effect
of strong interactions on the underlying weak decay; (2) the
magnitudes of Cabibbo-Kobayashi-Maskawa (CKM) matrix 
elements; (3) the origin of CP violation; and (4) the
physics of a host of rare processes, including flavor-changing
neutral current decays, which can probe physics beyond the 
standard model.
               
In this review,~\footnote{Invited talk presented at the 28th 
International Conference on High Energy Physics, 25--31 July 1996,
Warsaw, Poland} I focus primarily on measurements that
help to shed light on the dynamics of 
heavy-flavor decays. A detailed understanding of
these processes
is essential for determining the magnitudes of CKM elements.
The physics of heavy-flavor decays has also proved to be
a fascinating subject in its own right, since 
new theoretical methods have been devised to exploit the large
bottom and charm quark masses, 
leading to some remarkable predictions
that can be tested by experiment.

Progressing from
the simplest to the most complicated, I describe leptonic, 
semileptonic, and hadronic decays, including certain rare modes.
The most interesting experimental results on leptonic decays
are $D_s^+$ measurements, where leptonic decay is Cabibbo 
allowed and signals have been observed. Due to 
constraints on the length of this review, however, 
my discussions of 
semileptonic and hadronic decays are restricted almost entirely to 
decays of bottom mesons, omitting the vast amount
of important work on charm hadrons and bottom baryons. The exceptions
are the decays $D^0\to K^-\pi^+$ and $D^+_s\to\phi\pi^+$, which
I include because of their crucial role as normalizing modes in
both charm and bottom physics. 
From the perspective 
of dynamics, semileptonic
$B$ decays are especially interesting. 
Here, strong interaction effects are quite
important, but they are sufficiently simple in many cases
to allow
detailed theoretical predictions that can be tested experimentally.
Hadronic modes are the most difficult to describe, since strong
interactions affect both currents, and final-state interactions can
also come into play. However,
factorization has proved to be a useful
simplifying framework, 
at least in certain $b\to c$ decays with large energy release, 
and I review some of the measurements that test this idea. 
I also summarize new results on hadronic rare decays and 
discuss the question of penguin contributions 
to final states that are eigenstates of CP,
which would complicate the interpretation of CP 
violation measurements. 
Finally, I review the status of $b$-hadron lifetime 
measurements, which also have important implications for
our understanding of decay dynamics. 

The use of
semileptonic $B$ decays and 
$\bzbzbar$ oscillations to extract CKM elements is 
covered at this conference by Lawrence Gibbons,~\cite{Gibbons_talk}
who also discusses the important recent observations of
$b\to u\ell^-\overline\nu$ modes by CLEO.
Processes involving flavor-changing neutral currents are
reviewed by Andrzej Buras.~\cite{Buras_talk} 
Many of the
theoretical issues related to heavy-flavor dynamics are discussed
by Guido Martinelli.~\cite{Martinelli_talk} 
Rolf Landua~\cite{Landua_talk} discusses the spectroscopy of
heavy-flavor hadrons.
In addition, 
several recent articles review 
experimental~\cite{RichBurch_rev,Browder_rev}
and theoretical~\cite{Neubert_rev,Ali_rev,Bigi_rev,Neubert_BandCP} aspects of
heavy-flavor physics.

\section{Leptonic Decays}

In the leptonic decay of a charged meson $M$ (also of mass $M$) 
the quark and antiquark annihilate into 
a virtual $W,$ which then produces a charged lepton and
a neutrino. In this decay, the effect of strong interactions
can be parametrized by a single
``decay constant,'' $f_M^2\propto |\psi(0)|^2/M,$ 
where $\psi(0)$ is the amplitude for the quarks to have zero separation.
For a pseudoscalar
meson, the only available four-vector that can appear in the hadronic
current is $q^{\mu}$, the four-momentum of the meson. The
hadronic current is therefore given by $<0|J^{\mu}|M>=iV_{qQ}\,f_M\,q^{\mu}$,
where $V_{qQ}$ is the appropriate CKM matrix element.
The leptonic width is given by
\begin{eqnarray}
\Gamma_{\rm leptonic}={G_F^2\over8\pi} |V_{qQ}|^2 f_M^2 
M m_\ell^2 \biggl(1-{m_\ell^2\over M^2}\biggr)^2,
\label{eq:leptonic_width}
\end{eqnarray}  
where $m_{\ell}$ is the lepton mass and
the factor $m_{\ell}^2$ is a consequence of helicity
suppresssion.

There is great interest in obtaining
accurate measurements of decay constants
partly because they can be
compared with lattice 
QCD calculations,~\cite{Martinelli_talk} which
are becoming more reliable. In addition, 
decay constants are needed
to extract certain CKM matrix elements.
For example, the $\bzbzbar$ mixing rate is determined by
$\Delta M\propto f_B^2{\cal B}_B|V_{td}|^2$,
where $f_B$ is the $B$ meson decay constant and
${\cal B}_B$ is the bag constant.

Experimental study of leptonic decays of heavy-flavor 
mesons has been difficult for two reasons. 
First, the leptonic width 
is small compared to the total width
(unlike the case in $K^+$ decays), since 
$f^2_M M\to{\rm constant}$ for large $M$, whereas
the total decay 
rate is proportional to $M^5$.
Second, the presence
of a neutrino in the final state makes reconstruction of the
signal and rejection of background more difficult. The
$B^+$ leptonic decay rate is CKM suppressed  ($\propto|V_{ub}|^2$),
putting it beyond the reach of current measurements.
$D_s^+$ leptonic decay, however, is CKM favored ($\propto|V_{cs}|^2$),
and it has been measured by several experiments.

\begin{figure*}[tbh]
\epsfxsize=3.00in
\hbox{\hfill\hskip+1.1in\epsffile{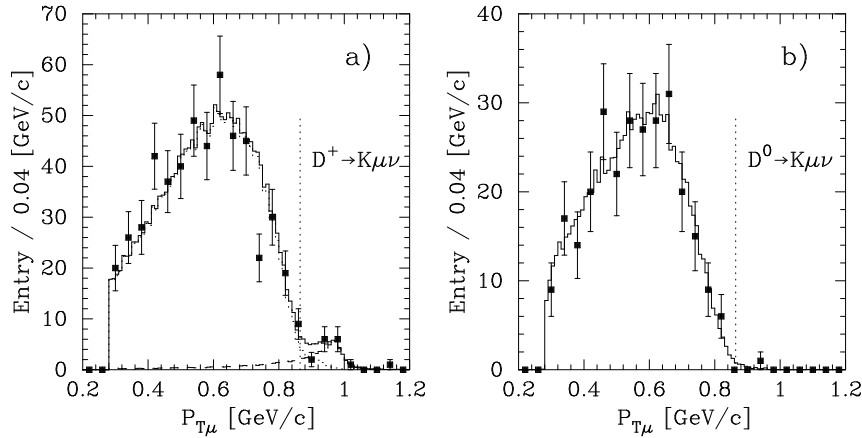}\hfill}
\vskip -10mm
\caption{
Observation of $\Dstomunu$ from E653: (a) The $P_{T\mu}$ distribution for
the sample of one-prong kinks, showing the excess beyond the endpoint
for the decay $D^+\to\bar K^0\mu^+\nu_{\mu}$ and (b) $P_{T\mu}$ distribution
for the sample of two-prong vees.\hfill}
\label{fig:E653_data}
\end{figure*}

The most recent measurement of $\Dstomunu$ has 
been reported by Fermilab E653,~\cite{E653_leptonic} in
which a 600 GeV/$c$ $\pi^-$ beam is incident on an active emulsion target.
Downstream of the target are 18 planes of silicon-strip detectors,
a magnetic spectrometer, and a muon system. 
The signature for $\Dstomunu$ is a muon track with 
a kink and large transverse momentum, $P_{T\mu},$ with respect to
the parent particle direction. 
Figure~\ref{fig:E653_data}(a) shows the $P_{T\mu}$ distribution for the
sample of one-prong kinks; most of the events are due to the decay
$D^+\to \bar K^0\mu^+\nu_{\mu},$ but there is a significant excess beyond
the kinematic endpoint, which is attributed to $\Dstomunu$ decays.
A fit to the $P_{T\mu}$ spectrum yields $23\pm6$ $\Dstomunu$ events.
As a check, E653 studies a sample of neutral, two-prong 
vee events (Fig.~\ref{fig:E653_data}(b)). 
This histogram shows a distribution of leptons mainly 
from $D^0\to K^-\mu^+\nu_{\mu},$
but there is no excess beyond the endpoint because the $D^0$ cannot
decay leptonically. 

\begin{table}[tb]
\begin{center}
\caption{Measurements of $\Dstomunu$. The branching fractions 
from WA75, CLEO, and E653 have
been updated to reflect the 1996 PDG value for $B(D_s^+\to\phi\pi^+)$.
The normalization of the WA75 experiment is somewhat complicated,
since the relative $D^0$ to $D_s^+$ cross section is determined by
a different experiment (ACCMOR).
The BES signal includes one $D_s^+\to\tau^+\nu_{\tau}$ candidate.
\hspace*{\fill}}\label{tab:dstomunu}
\vspace{0.4cm}
{\scriptsize
\begin{tabular}{|lccc|}\hline
Expt & Norm. Mode & Events & $B(D_s\to\mu^+\nu_{\mu})$ \\   
     &            &        &  \qquad     $/10^{-3}$    \\ \hline
WA75                 & $D^0\to\mu^+\nu_{\mu}X$     & $8.5^{+3.8}_{-3.0}$ &
$4.1\pm 1.6\pm2.2$ \\
CLEO                 & $D_s\to\phi\pi^+$           & $47\pm10$ & 
$6.6\pm1.9\pm1.7$\\
E653                 & $D_s\to\phi\mu^+\nu_{\mu}$  & $23\pm6$ &
$3.0\pm1.3\pm0.8$\\    
BES                  & absolute                    & 3  & 
$15^{+13+\ 3}_{-6-2}$   \\   
Average & & & $4.6\pm0.8\pm1.2$\\
\hline
\end{tabular}
}
\end{center}
\end{table}

The status of $\Dstomunu$ results is summarized 
in Table~\ref{tab:dstomunu}.
In addition to the E653 measurement, there are
previous results from WA75,~\cite{WA75_leptonic} 
CLEO II,~\cite{CLEOII_leptonic} and BES.~\cite{BES_leptonic}
Normalization of the signal within each experiment is an important issue.
As I will discuss later, the absolute scale of the $D_s^+$ branching
fractions is not well known. I have renormalized all but the
BES result (which is an absolute measurement) to 
the PDG 96~\cite{PDG96} value 
$B(D_s^+\to\phi\pi^+)=(3.6\pm0.9)\%.$ 
Using the renormalized branching fractions, I obtain the world average
branching fraction $B(\Dstomunu)=(4.6\pm0.8\pm1.2)\times 10^{-3},$ which
corresponds to the decay constant value 
\begin{eqnarray}
f_{D_s}=(241\pm21\pm30)\ {\rm MeV},
\end{eqnarray}
where the first error combines the statistical and systematic uncertainties
and the second error is due to the normalization. There are many lattice
QCD calculations 
of $f_{D_s}.$ A recent 
(preliminary) result from the MILC
collaboration~\cite{ds_lattice} is
$f_{D_s}=(211\pm 7\pm 25\pm 11)\ {\rm MeV},$
where the last error is the estimated uncertainty due
to the quenched approximation.

Also reported at this conference is
a new measurement of $\Dstotaunu$ from L3.~\cite{L3_leptonic} In this mode,
the helicity suppression is nearly absent because of the large $\tau$-lepton
mass.
L3 reports a preliminary value of 
$B(\Dstotaunu)=(8.9\pm2.6\pm1.1\pm2.1)\%$, which corresponds to 
$f_{D_s}=(351\pm53\pm19\pm37)\ {\rm MeV}.$ Averaging this result
with those from $\Dstomunu$ gives 
$f_{D_s}=(255\pm20\pm31)\ {\rm MeV}.$ 

Although data samples are much too small to observe 
$B^+\to\tau^+\nu_{\tau}$ at the expected branching fraction
(roughly $0.5\times 10^{-4}$ to $1.0\times 10^{-4}$), an
anomalously large rate could arise as the
result of 
physics beyond the standard model. 
ALEPH,~\cite{ALEPH_btotaunu} CLEO II,~\cite{CLEOII_btotaunu} 
and L3~\cite{L3_leptonic} have obtained 
upper limits for this mode
with a typical value of 
$B(B^+\to\tau^+\nu_{\tau})< 2\times 10^{-3}$ (90\% C.L.).
In the process $B^-\to\ell^-\bar\nu_{\ell}\gamma$,
($\ell=e,\ \mu$), the helicity
suppression is removed and the branching fractions for
$e$ and $\mu$ are expected to be nearly the same, 
in the range $1.0\times 10^{-6}$ to $4.0\times 10^{-6}$.
These model-dependent predictions can be compared with new
CLEO limits~\cite{CLEO_btomunugamma}: 
$B(B^-\to \mu^-\bar\nu_{\mu}\gamma)<5.2\times 10^{-5}$ and
$B(B^-\to e^-\bar\nu_{e}\gamma)<2.0\times 10^{-4}$ at 90\% C.L.

\section{Semileptonic Decays}
A vast amount of information has been obtained on 
semileptonic decays of heavy flavors.   
I will begin with a simple, physical picture 
of semileptonic 
decay dynamics~\cite{RichBurch_rev} and then 
turn to measurements of 
exclusive decays, including detailed
studies of form factors. Finally, I discuss
the inclusive semileptonic branching fraction
and its implications.
\subsection{Dynamics of Semileptonic Decays}

Semileptonic decays, because of their simplicity,
provide an excellent laboratory in which to study
the effect of nonperturbative QCD interactons 
on the weak decay process.
The matrix element can be written as
the product of a leptonic current, which is exactly known, and
a hadronic current, which can be parametrized in terms of form
factors. The form factors are Lorentz-invariant functions that may
be expressed in terms of $\qsq,$ the square of the mass of the
virtual $W$.

Because semileptonic decays 
produce at least three final-state particles, $q^2$ is
a variable that ranges from $q^2_{\rm min}=m_{\ell}^2$ 
(which is nearly
zero for $\ell=e$ or $\ell=\mu$) to a maximum 
value $q^2_{\rm max}=(m_B-m_X)^2,$ where $X$ is the final-state
hadron or hadronic system. The variation of the amplitude with
$\qsq$ is of great interest, since it probes the effects
of strong interactions on the decay. In fact, $\qsq$ determines
the recoil velocity of the daughter hadron in the $B$ rest frame:
\begin{eqnarray}
w=\gamma_X={{E_X}\over m_X}=
{M_B^2+m_X^2-\qsq\over 2M_B m_X},
\end{eqnarray}
which also shows that by measuring $E_X$ in the
$B$ rest frame, one can determine $\qsq$.
This result is simply the two-body decay formula applied to
a situation in which one of the particles has the
variable mass $\sqrt{q^2}$.
 
\begin{figure}[tbh]
\epsfxsize=2.0in
\hbox{\hfill\hskip+0.5in\epsffile{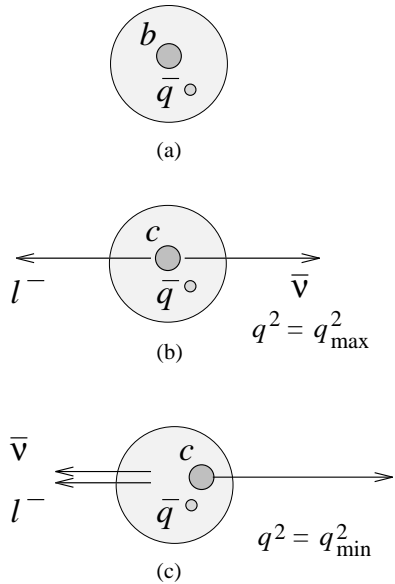}\hfill}
\vskip 3mm
\caption{Kinematic configurations in semileptonic $B$ decay
(a) $B$ meson before decay; (b) configuration 
at $q^2_{\rm max}\ (w=1)$,
where the daughter $c$ quark has little velocity relative to the
spectator; and (c) configuration at 
$q^2_{\rm min}\approx 0$ ($w=w_{\rm max}$),
where the daughter $c$ quark has large velocity.\hspace*{\fill}}
\label{fig:sl_configs}
\end{figure}

Figure~\ref{fig:sl_configs} shows a $B$ meson before decay and
two extreme decay configurations.
At the largest value of
$\qsq$ ($w=1$), known as the zero-recoil configuration, 
the full energy of the $B$ goes into the
masses of the daughter hadron and the $W,$ 
as shown in Figure~\ref{fig:sl_configs}(b).
The daughter hadron and the $W$ are therefore produced at rest 
with respect to the $B$
meson, and the lepton and neutrino are back to back. 
If both the initial and the daughter quarks are very
heavy, the hadronic system is nearly undisturbed 
for configurations at or near $q^2_{\rm max}$: one static source
of a color field is simply replaced by another. Color magnetic moment
effects, which are proportional to $1/m_Q,$ are absent in this
limit, so the initial and final quarks are completely equivalent. 
The overlap between the initial and final hadron 
wave functions is therefore very large, leading to a small
uncertainty in the transition form factor and consequently
reliable predictions for the rate as a function of $|V_{cb}|$.
This ideal situation near $q^2_{\rm max}$ appears to 
apply reasonably well to $b\to c\ell^-\bar\nu$ but 
not to $b\to u\ell^-\bar\nu$ decays, where only one quark is heavy.

As $q^2$ decreases (and $w$ increases), 
the lepton and neutrino
become more collinear, and the daughter quark recoils at
higher and higher velocity with respect to the spectator
(Fig.~\ref{fig:sl_configs}(c)). 
The rapidly moving daughter quark must exchange gluons 
with the spectator quark in order to form a bound state. 
The faster the daughter quark, the more this gluon
exchange suppresses the form factors, and hence the
amplitude. This interaction is nonperturbative, and methods
such as lattice QCD or QCD sum rules have been used to
calculate the $q^2$ dependence of the form factors. 
In general, a larger range of recoil velocities leads to a 
larger falloff in the decay form factors. 
For $\btodstlnu$ the range is a modest $\Delta w=0.5,$
whereas for $\btopilnu$ the range is $\Delta w=17.9,$ so the pion
becomes very relativistic. Thus, $\btoclnu$ decays  
are more tractable theoretically, both because they have a fairly
reliable normalization point at $q^2_{\rm max}$ and because the range
of recoil velocities is relatively small. 
 
To understand studies of semileptonic decay dynamics,
it is important to know that for a decay
of the form $P\to P'\ell^-\bar\nu$, where both $P$ and $P'$ are
pseudoscalar mesons, there is only one operative form factor, 
$F(q^2),$ assuming that 
the mass of the lepton is neglected. (This approximation is
very good for $\ell^-=e^-$ or $\ell^-=\mu^-$.) However, for the case
$P\to V\ell^-\bar\nu$, where $V$ is a vector meson, the spin-polarization
vector of $V$ allows one to construct additional terms in the
hadronic current, and there are three form factors, $A_1(\qsq)$,
$A_2(\qsq)$, and $V(\qsq)$, that are operative when the mass of the
lepton is neglected. 

Our understanding of $b\to c$ semileptonic decay has improved dramatically
with the development of heavy-quark effective 
theory~\cite{Isgur_Wise_1,Isgur_Wise_2,Neubert_rev} (HQET). In
the heavy-quark symmetry limit ($m_b\to\infty$ and $m_c\to\infty$),
all of the form factors discussed above are related to a single
form factor, the Isgur-Wise funtion $\xi(w)$:
\begin{eqnarray}
F(\qsq)&=&{(M+m_{P'})\over 2\sqrt{Mm_{P'}}}\xi(w)\nonumber\\
V(\qsq)&=&A_2(\qsq)={A_1(\qsq)\over \bigl[1-{\qsq\over (M+m_V)^2}\bigr]}
\nonumber\\
&=&{(M+m_{V})\over 2\sqrt{Mm_{V}}}\xi(w).
\label{eq:hqs}
\end{eqnarray} 
These symmetry relations represent a major simplification, even though
they do not tell us the form of $\xi(w)$. However, in the heavy-quark
symmetry limit, there is one additional result, $\xi(1)=1$, which
is the form factor normalization in the 
zero-recoil configuration.

In the real world, of course, the quark masses are not infinite,
so these results cannot be exact.
The heavy-quark symmetry limit is thus only the first term in
the HQET expansion in $1/m_{Q}.$ 
In particular, the simple relations among the semileptonic
decay form factors given above are somewhat 
modified.~\cite{Neubert_rev,Close_and_Wambach,Neubert_BandCP}
Measurements of the kinematic distributions in semileptonic decay 
provide an important check of these HQET-based predictions.
For $\btoulnu$ decays,
HQET is not directly applicable, but 
lattice QCD calculations are beginning to produce useful predictions.

\subsection{Exclusive Semileptonic $B$ Decays}

Many new results were reported at this conference on 
$\bar B\to D\ell^-\bar\nu$ and
$\bar B\to D^*\ell^-\bar\nu$, 
which account for roughly
two-thirds of the inclusive semileptonic rate,
and on $\bar B\to D^{**}\ell^-\bar\nu$, where
$D^{**}$ indicates an orbitally excited charm meson.
Improvements in $\btodlnu$ measurements have come 
rather slowly, partly because $\btodlnu$ has
a substantial feed-down background from $\btodstlnu$.
But experimenters have also focused more on $\btodstlnu$
because it is the preferred mode for measuring $|V_{cb}|$.
In $\bar B\to D^*\ell^-\bar\nu,$ 
there are no leading order ($1/m_Q$) power 
corrections to the form factor
normalization at high $\qsq$ (Luke's theorem~\cite{Lukes_theorem}), 
where the $|V_{cb}|$ 
measurement is performed. This result does not hold for 
$\bar B\to D\ell^-\bar\nu,$ which has a further (but related) problem:
its rate is highly suppressed at high $\qsq$
by the kinematic factor $p_D^3,$ which
arises from the $p$-wave nature of this decay. Nevertheless,
it is very important for testing HQET to study the form factor shape for
$\bar B\to D\ell^-\bar\nu$ and to compare it with the form factor shapes
in $\bar B\to D^*\ell^-\bar\nu.$ Measurements of both 
$\bar B\to D\ell^-\bar\nu$ and $\bar B\to D^*\ell^-\bar\nu$ also
provide information that can be used to test whether
factorization is a good description of the modes 
$\bar B\to D^{(*)}h$, where $h$ is a hadron.

\begin{figure}[tbh]
\epsfxsize=2.9in
\hbox{\hfill
\epsffile{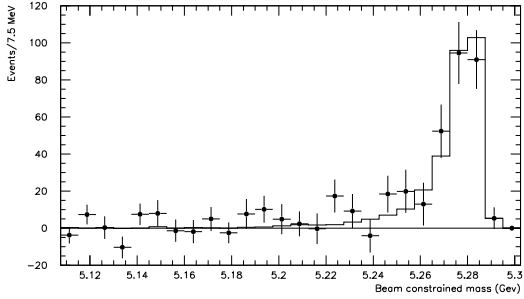}\hfill}
\vskip -4mm
\caption{CLEO II measurement of $B(\bar B^0\to D^+\ell^-\bar\nu)$
using ``neutrino reconstruction.''
The points with error bars are the 
background-subtracted distribution of 
$M(D\ell^-\bar\nu)$, where the neutrino four-momentum is 
estimated from the missing momentum in the event.\hspace*{\fill}}
\label{fig:btodlnu_mb}
\end{figure}

\begin{figure}[tbh]
\epsfxsize=3.0in
\hbox{\hfill
\epsffile{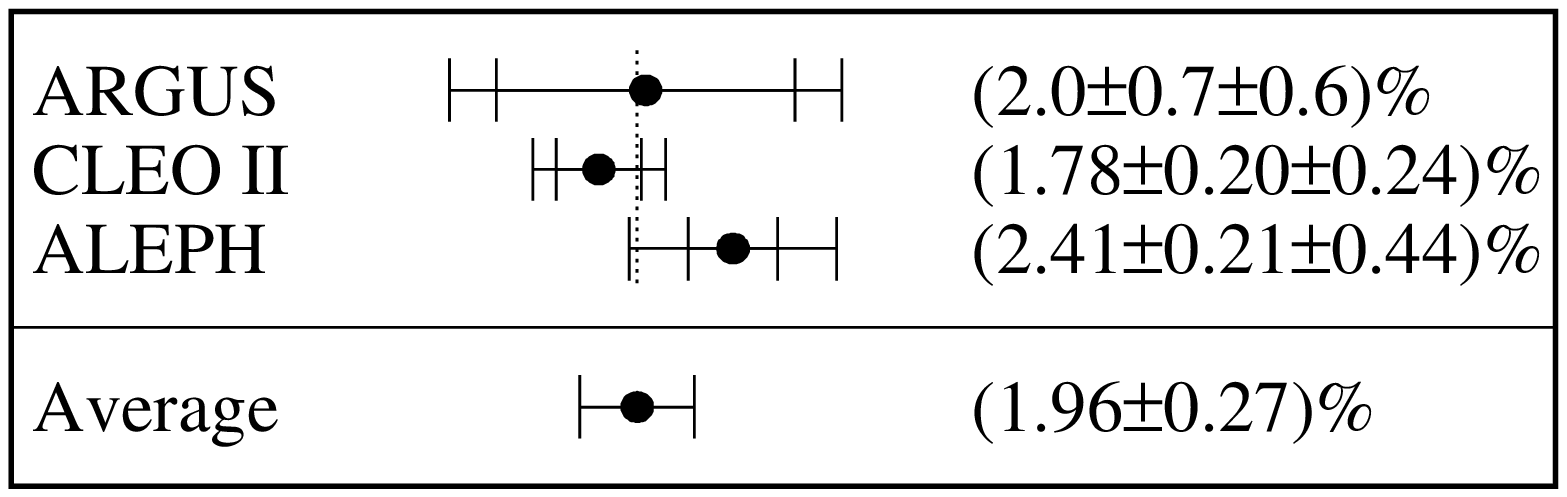}\hfill}
\vskip -4mm
\caption{Summary of measurements of the branching fraction for
$\bztodlnu$. In the values listed, the normalization 
is based on the 1994 PDG value for $B(\dtokpi),$ which sets the
scale for $D^+$ branching fractions. Updating the measurements
to reflect the new value for $B(\dtokpi)$ used in this paper, one
obtains $B(\bztodlnu)=(2.03\pm0.28)\%$.   
\hspace*{\fill}
} 
\label{fig:btodlnu_br}
\end{figure}

\begin{figure}[tbh]
\epsfxsize=2.9in
\hbox{\hfill\hskip+0.1in
\epsffile{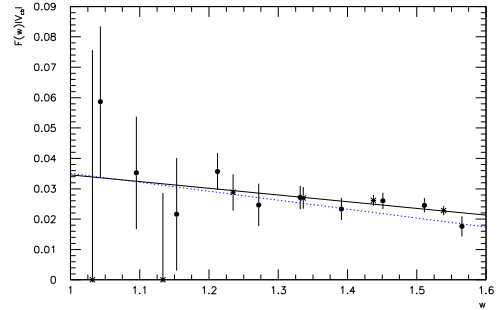}\hfill}
\vskip -4mm
\caption{
CLEO II measurement of the function $V_{cb}F(w)$ 
for $\bztodlnu$. The points and asterisks represent the data
from the neutrino-reconstruction and missing-mass methods,
respectively. The solid line is the average of the form
factor slopes from these two methods, and the dotted line has
the slope $\hat\rho^2$ as obtained from the CLEO II measurement of
$|V_{cb}|$ using $\btodstlnu.$ \hspace*{\fill}}
\label{fig:btodlnu3}
\end{figure}

Three recent measurements, two from CLEO II~\cite{CLEO_btodlnu} 
and one from
ALEPH~\cite{ALEPH_btodstlnu}, have significantly improved the situation for 
$\bar B^0\to D^+\ell^-\bar\nu.$ This mode typically has less
background than $B^-\to D^0\ell^-\bar\nu,$ since all $D^{*0}$
and about 70\% of $D^{*+}$ decays produce $D^0$ background.
One of the CLEO analyses uses a novel
``neutrino reconstruction'' technique, which relies on the 
near hermiticity of the detector. (This method was also used in 
the CLEO II measurements~\cite{CLEO_btorholnu} 
of $\bar B\to\pi\ell^-\bar\nu$ and 
$\bar B\to\rho\ell^-\bar\nu.$) The four-momentum
of the neutrino is determined from the missing momentum vector of the
event; once this quantity is obtained, a beam-energy constrained 
$B$ mass peak is reconstructed, just as for a hadronic decay analysis. 
Although this idea may not seem new in itself, the real 
advance lies in
the method for obtaining good resolution on the neutrino momentum.
The main idea is to restrict the sample to $B\bar B$ events 
in which there is only
one semileptonic decay (by requiring no additional leptons) and 
no $K_L$'s (by requiring a small missing mass in the event).
Together, these and other requirements lead to a resolution
on the neutrino energy in signal events of about 110 MeV. 
Figure~\ref{fig:btodlnu_mb} shows the background-subtracted
distribution of $M(D^+\ell^-\bar\nu_{\rm miss}),$ where $\nu_{\rm miss}$
refers to the neutrino as reconstructed from the missing momentum
vector. The signal contains $238\pm 27$ events. An alternative
CLEO analysis, which uses the more traditional approach of
analyzing the missing mass recoiling against
the $D^+\ell^-$ system, obtains a much higher yield at the expense
of more background. The average of the branching fractions 
from these two analyses is presented in Fig.~\ref{fig:btodlnu_br}. 
The table also includes earlier results from ARGUS~\cite{ARGUS_btodlnu} 
and the new
measurement from ALEPH~\cite{ALEPH_btodstlnu}, 
which has a signal of $266\pm24$ events.
The branching fractions reported by the experiments 
assume $D^+$ branching fractions
from the 1994 Particle Data Book~\cite{PDG94}. 
I have computed two averages, one
from those values and a second using values  
updated to reflect the new $D^0\to K^-\pi^+$  
branching fraction presented in this paper.  
\vskip -3mm
Figure~\ref{fig:btodlnu3} shows the measured values of the quantity 
$F(w)V_{cb}$ as a function of $w$ for the two CLEO analyses. 
The quantity $F(w)V_{cb}$ is only a part of the decay amplitude:
the $w$ dependence due to $p$-wave kinematics 
has been factored out, allowing us to see the form factor
itself. It is apparent from Fig.~\ref{fig:btodlnu3} that the dependence
of the form factor on $w$ is mild, and it is well described by a 
linear function of slope $\rho^2$. The CLEO II~\cite{CLEO_btodlnu}
measurements give $\rho^2=0.64\pm0.18\pm0.10,$ while 
ALEPH~\cite{ALEPH_btodstlnu}
obtains $\rho^2=0.00\pm0.49\pm0.38$.
(Both results are preliminary.) 
The figure also shows
the linear function extracted from the CLEO $\btodstlnu$ data,
whose slope is consistent with that from $\btodlnu$.
Below I will present a systematic comparison of the various measurements
of the form factor slopes.

\begin{figure}[tbh]
\epsfxsize=2.9in
\hbox{\hfill\hskip+0.1in
\epsffile{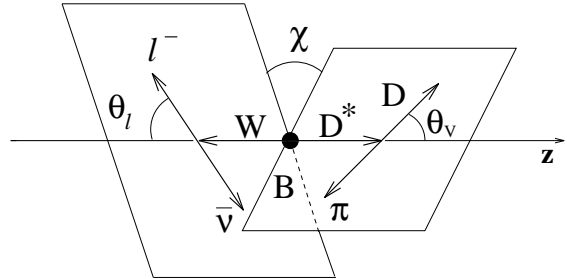}\hfill}
\vskip 1mm
\caption{Decay angles used in the CLEO II 
measurement of the $\btodstlnu$ form factors. The angle
$\theta_{\ell}$ is measured in the $W$ rest frame,
while $\theta_V$ is measured in the $D^*$ rest frame. 
The azimuthal angle $\chi$ is measured between the $W$ and $D^*$
decay planes.\hspace*{\fill}}
\label{fig:btodstlnu_angles}
\end{figure}

What measurements allow us to test the predictions of 
HQET? With data from semileptonic
decay, there are essentially two approaches: (1) comparison of the
different form factors governing $\btodstlnu$ with each other
and (2) comparison of $\btodlnu$ with $\btodstlnu.$ 
Figure~\ref{fig:btodstlnu_angles} shows the decay angles that,
together with $q^2,$ are used in the CLEO II measurement of the
$\btodstlnu$ form factors.~\cite{CLEO_btodstlnuff_prl,CLEO_btodstlnuff_pre} 
The form factors are measured by performing a joint, four-dimensional
maximum likelihood fit to these variables.
The analysis is performed in two modes,
$\bztodstlnu$, which is fairly clean 
($779\pm42$ signal events over $161\pm28$ background
events), and $\bmtodstlnu,$ which is somewhat more difficult 
($417\pm33$ signal events over a background of $270\pm21$ events).  

\begin{figure}[tbh]
\epsfxsize=2.9in
\hbox{\hfill
\epsffile{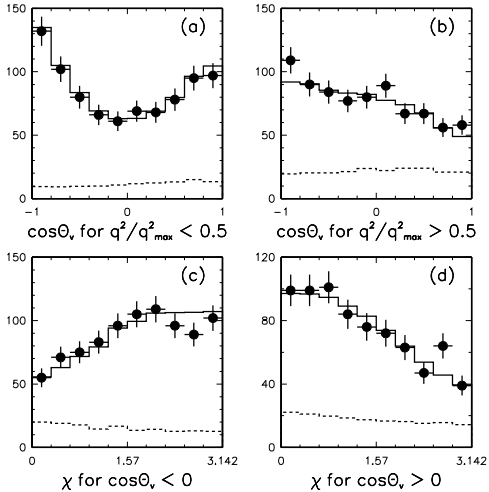}\hfill}
\vskip 1mm
\caption{Distributions of $\cos\theta_V$ and $\chi$ in
the CLEO II measurement of the $\btodstlnu$ form factors. 
The
histograms show (a) $\cos\theta_V$ for the lower half of the
$q^2$ range, (b) $\cos\theta_V$ for the upper half of the $q^2$ range,
(c) $\chi$ for the lower half of the $\cos\theta_V$ range, and
(d) $\chi$ for the upper half of the $\cos\theta_V$ range. 
The points with error bars are the data; the solid histogram is the
fit, including backgrounds; and the dashed line is the contribution
from backgrounds. The
interpretation of these histograms is discussed in the text.
\hspace*{\fill}}
\label{fig:btodstlnu_dists}
\end{figure}

Figure~\ref{fig:btodstlnu_dists} shows some of the 
fit projections
for the combined $\btodstlnu$ modes. The two upper histograms
compare the distributions of $\cos\theta_V$ in the lower and
upper half of the $q^2$ range. Although acceptance effects, which
are taken into account in the fit, 
gradually reduce the efficiency as $\cos\theta_V$ increases,
it is apparent that in the lower
$q^2$ range there is a strong forward-backward-peaking component.
At low $q^2$, the lepton and antineutrino become collinear, with zero
net spin along their common direction, forcing the $D^*$ also to have 
zero helicity. 
This effect produces a distribution
$dN/d\cos\theta_V\propto\cos^2\theta_V$.
In contrast, at very high $q^2$, the $D^*$ is 
nearly at rest and unpolarized,
producing a $\cos\theta_V$ distribution that is uniform, apart from
acceptance effects.
The lower two histograms show distributions of the azimuthal angle
$\chi$ for the lower and upper range of $\cos\theta_V.$ 
The correlation between these histograms arises from a quantum
interference term proportional to the difference between negative
and positive helicity amplitudes, multiplied by the zero 
helicity amplitude. The upward
slope of the first histogram (c) and the downward slope of the
second histogram (d) are allowed by $(V-A)(V-A)$ or $(V+A)(V+A),$
but this correlation is forbidden by a mixed coupling $(V\mp A)(V\pm A)$. 

In the framework of HQET, it is useful to construct
quantities that have simple behavior in the heavy-quark
symmetry limit. It is conventional to define~\cite{Neubert_rev} 
the form-factor ratios
\begin{eqnarray}
R_1(w)\equiv 
\biggl[1-{q^2\over(m_B+m_{D^*})^2}\biggr]{V(q^2)\over A_1(q^2)}\nonumber\\ 
R_2(w)\equiv 
\biggl[1-{q^2\over(m_B+m_{D^*})^2}\biggr]{A_2(q^2)\over A_1(q^2)}
\end{eqnarray}
and the HQET version of the $A_1$ form factor, called $h_{A_1}$:
\begin{eqnarray}
h_{A_1}(w)={2\sqrt{m_Bm_{D^*}}\over (m_B+m_{D^*})}{A_1(q^2)\over
\biggl[1-{q^2\over (m_B+m_{D^*})^2}\biggr]}.
\label{eq:r1r2}
\end{eqnarray}
(The $A_1$ form factor is singled out because it contributes to
all three helicity amplitudes and
dominates the rate at high $q^2$.)
In the heavy-quark symmetry limit, $R_1(w)\to 1$ and $R_2(w)\to 1$,
and $h_{A_1}(w)\to \xi(w)$, as can be seen from Eq.~\ref{eq:hqs}.
The fit determines $R_1(w=1)$ and $R_2(w=1)$, as well as
the dimensionless form factor slope 
$\rho^2_{A_1}$ of $h_{A_1}.$ (Since the true form of $h_{A_1}(w)$ is 
not known, several different functions are tried in addition 
to a simple linear function with slope $\rho^2_{A_1}.$
The details of these alternative fits are discussed in the 
references.~\cite{CLEO_btodstlnuff_prl,CLEO_btodstlnuff_pre})
Table~\ref{tab:ff_results}
compares the measured values 
for $R_1(1)$ and $R_2(1)$ 
obtained from the combined fit to $\bztodstlnu$ and $\bmtodstlnu$
with theoretical predictions from 
Neubert,~\cite{Neubert_rev} Close and
Wambach,~\cite{Close_and_Wambach} and the 
ISGW2 model.~\cite{ISGW2}
Within errors, the measurements are in good agreement with the predictions, 
and they follow the expected pattern $R_1(1)>1$ and $R_2(1)<1.$ 
(CLEO has published the fit results for $\bztodstlnu$ only; the values in
Table~\ref{tab:ff_results} are obtained from a combined fit
to both $\bztodstlnu$ and $\bmtodstlnu.$ 
These results are preliminary and are described in a paper 
submitted to this conference.)

The slope $\rho^2_{A_1}$ describes nonperturbative QCD physics
and can only be calculated with methods such as lattice QCD or QCD
sum rules, with typical values ranging from 0.5 to 1.0. 
It is closely related to another parameter, 
$\hat\rho^2,$ which is extracted together with $|V_{cb}|$
in studies of $\btodstlnu$. 
Such analyses, which use only the $q^2$ distribution, 
have been performed by ARGUS, CLEO, and the LEP experiments.
Because the three form factors cannot be separated using the
$q^2$ distribution only, $\hat\rho^2$ is the slope of a function
${\cal F}(w)$ that has 
complicated dependence on all three form factors. Neubert gives the
relation~\cite{Neubert_rhosqrelation} 
$\hat\rho^2=\rho^2_{A_1}-f(R_1,R_2),$ where the function
$f(R_1,R_2)\approx0.2$ for the values of $R_1$ and $R_2$ quoted above.

\begin{table}[tb]
\begin{center}
\caption{CLEO II measurements of form factor ratios for $\btodstlnu$
and comparison with theoretical predictions. 
The CLEO results are based on a combined fit to both $\bztodstlnu$ and
$\bmtodstlnu$ and are preliminary.\hspace*{\fill}}
\label{tab:ff_results}
\vspace{0.4cm}
{\scriptsize
\begin{tabular}{|lcc|}\hline
           & $R_1(w=1)$ & $R_2(w=1)$  \\ \hline
CLEO II    & $1.24\pm0.26\pm0.12$ & $0.72\pm0.18\pm0.07$ \\
Neubert    & $1.3\pm0.1$          & $0.8\pm0.2$ \\
Close \&      & 1.15           & 0.91 \\
\ \ Wambach   &                &      \\
ISGW2      & 1.27                 & 1.01 \\
\hline    
\end{tabular}
}
\end{center}
\end{table}

\begin{figure}[tbh]
\epsfxsize=2.9in
\hbox{\hfill
\epsffile{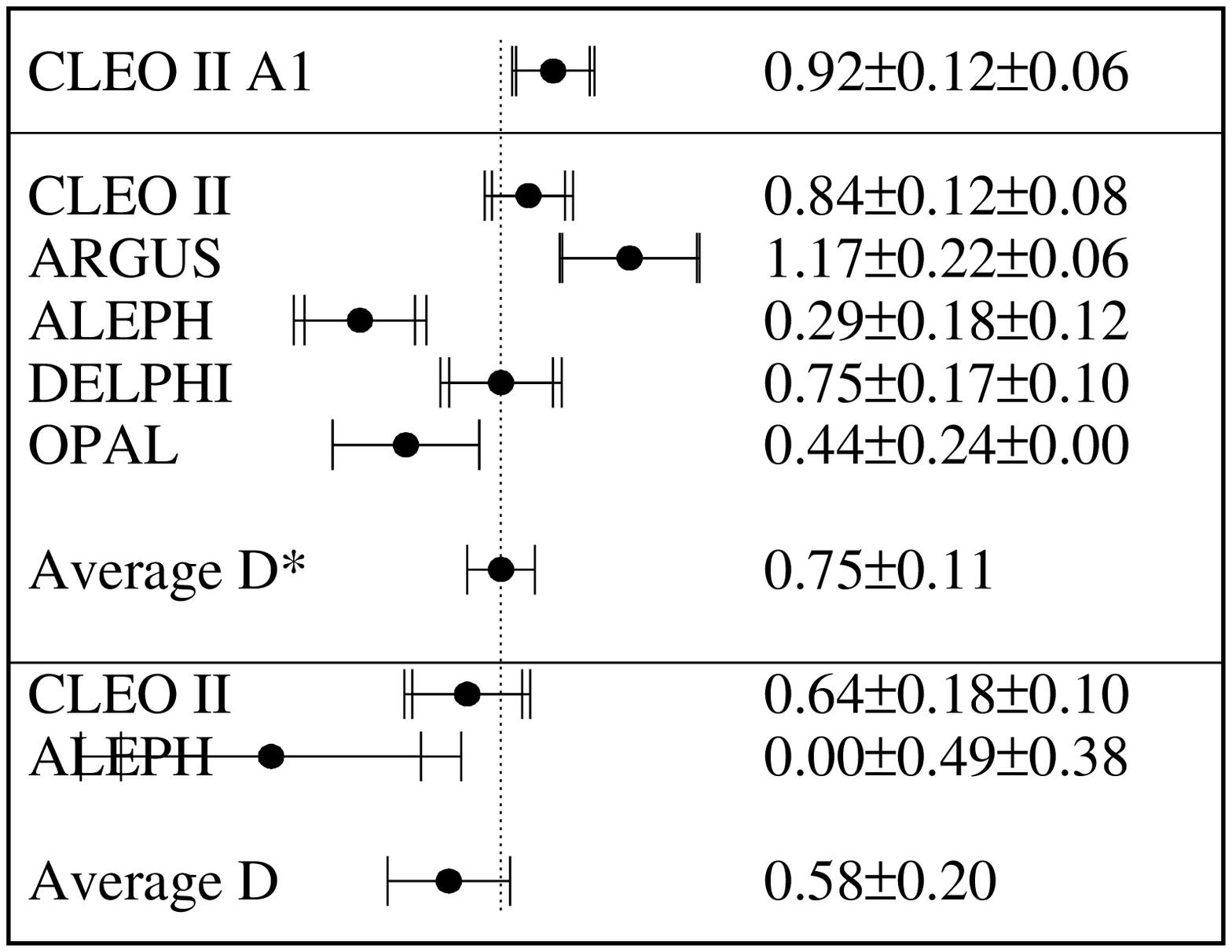}\hfill}
\vskip -4mm
\caption{Measurements of form factor slopes in $\btodstlnu$
and $\btodlnu$. The top measurement
is $\rho^2_{A_1}$ from the CLEO II four-dimensional fit to $\btodstlnu$. 
Below it are measurements of $\hat\rho^2$ from $\btodstlnu$, which are
expected to be slightly lower than $\rho^2_{A_1}.$ 
The vertical
dotted line corresponds to the average value of $\hat\rho^2$. 
The bottom group (CLEO II, ALEPH) is the slope measured in $\btodlnu$. 
\hspace*{\fill}}
\label{fig:rhosq}
\end{figure}

Figure~\ref{fig:rhosq} lists the form factor slopes for both
$\btodlnu$ and $\btodstlnu$. The first measurement is $\rho^2_{A_1}$
from the CLEO II four-dimensional fit to the $\btodstlnu$ kinematic 
distributions~\cite{CLEO_btodstlnuff_pre}; 
the second group of measurements is $\hat\rho^2$ from
the $q^2$ distribution of $\btodstlnu$; and the third group is
the slope of the form factor for $\btodlnu$. The average value of the
second group, which consists of measurements from
CLEO,~\cite{CLEO_rhosq} ARGUS,~\cite{ARGUS_rhosq}
ALEPH,~\cite{ALEPH_btodstlnu} DELPHI,~\cite{DELPHI_rhosq} and
OPAL~\cite{OPAL_rhosq}) is
$\hat\rho^2=0.75\pm0.11$, where the slightly inflated error
takes into account the somewhat poor agreement among the
measurements. As expected, this value is slightly lower than 
$\rho^2_{A_1}$, although the uncertainties are still
too large to draw a strong conclusion. 
Furthermore, the slope for $\btodlnu$ is consistent
with those measured for $\btodstlnu$, 
although here again improvements in the 
uncertainties are very desirable. 

In summary, we can say that $R_1(1)$ and $R_2(1)$ are quite consistent
with HQET predictions (they are actually consistent with 
the heavy quark symmetry limit $R_1=R_2=1$ itself), 
as are the relative sizes of form factor
slopes for $\btodlnu$ and $\btodstlnu$. With larger data samples, 
we can expect substantial improvements in these measurements.

Large values of $\rho^2_{A_1}$ (or $\hat\rho^2$) have an
interesting consequence:
the rates for $\btodlnu$ and $\btodstlnu$ fall off
faster with increasing recoil velocity, and 
the rates for $\btodststlnu$ modes increase. For example, Dunietz
gives the prediction~\cite{Dunietz_dtokpi}
\begin{eqnarray}
{B(\bar B\to D^{**}\ell^-\bar\nu)\over
B(\bar B\to X_c\ell^-\bar\nu)}
\approx 2(\rho^2-{1\over 4})(0.08\pm0.04).
\label{eq:btodststlnu}
\end{eqnarray}
New measurements of both $B(\bar B\to D_1\ell^-\bar\nu)$
and $B(\bar B\to D_2^*\ell^-\bar\nu)$
were submitted to this conference by ALEPH~\cite{ALEPH_btodststlnu} 
and CLEO.~\cite{CLEO_btodststlnu} 
Some of the results are listed in Table~\ref{tab:btodststlnu},
along with earlier results from OPAL.~\cite{OPAL_btodststlnu}
In general, such measurements make assumptions
regarding $D_1$ and $D_2^*$ branching fractions and
the absence of additional particles produced in the $B$ decay,
so some caution is advisable in interpreting the results.  
\begin{table}[tb]
\begin{center}
\caption{
Measurements of $\btodststlnu$ modes. The results marked
with an asterisk$^*$ were submitted to this conference and
are preliminary. The ALEPH $D_1\ell^-\bar\nu$ measurement is
an average over $D_1^+\ell^-\bar\nu$ and $D_1^0\ell^-\bar\nu.$
\label{tab:btodststlnu}
\hspace*{\fill}}
\vspace{0.2cm}
{\scriptsize
\begin{tabular}{|lcc|}\hline
Expt         & $B(B^-\to D_1^0\ell^-\bar\nu)$ &
$B(B^-\to D_2^{*0}\ell^-\bar\nu)$  \\ \hline
OPAL         & $(2.02\pm0.51\pm0.47)\%$ & $(0.88\pm0.35\pm0.18)\%$ \\
ALEPH$^*$    & $(0.74\pm0.16)\%$        & $<1\%$ (95\% C.L.) \\
CLEO II$^*$  & $(0.49\pm0.13\pm0.06)\%$ & $<1\%$ (95\% C.L.) \\
\hline    
\end{tabular}
}
\end{center}
\end{table}
\subsection{Inclusive Semileptonic Branching Fraction}
The inclusive semileptonic branching fraction is defined as
\begin{eqnarray}
B_{\rm SL}={\Gamma(\bar B\to Xe^-\bar\nu_e)\over
\sum_{\ell}\Gamma(\bar B\to X\ell^-\bar\nu)+
\Gamma_{\rm Had}+\Gamma_{\rm Rare}},
\label{eq:slbr}
\end{eqnarray}
where $\Gamma_{\rm Had}$ and $\Gamma_{\rm Rare}$ are the partial widths to
hadronic and rare final states.
Most measurements of $\slbr$ do not actually determine
the $b$ hadron species, because only the lepton is identified 
to keep the detection efficiency as high as possible. Thus,
most measurements of $\slbr$ at the $\Upsilon(4S)$ are an average over
the $\bar B^0$ and $B^-$ (with the notable exception of a new
CLEO~\cite{CLEO_slbrtagged} 
measurement submitted to this conference, which determines the
value of $\slbr$ for $\bar B^0$ mesons).
At the $Z$, the $B_s$ and $b$ baryons
are included in the samples as well. Since $b$ baryon lifetimes are typically 
shorter than those of $B$ mesons, and it is reasonable to assume that
the semileptonic rates are very similar, we expect the 
average $b$ hadron semileptonic branching fraction measured at the $Z$
to be slightly lower than that measured at the $\Upsilon(4S)$.
As a reminder of this small difference, I will use the symbol 
$\slbr^Z$ to refer to measurements performed at the $Z$. 

Precise measurements of $\slbr$ allow us
to determine the fraction of exclusive semileptonic modes 
that have been identified. 
%
For the purpose of comparing the inclusive with the 
sum of exclusive semileptonic 
$B$ branching fractions, I compute an average value of $\slbr$ using
the two measurements at the $\Upsilon(4S)$ that are based on the
dilepton method, which has very little model dependence.
Averaging the measurements from CLEO,~\cite{CLEO_dilepton} 
$\slbr=(10.49\pm0.17\pm0.43)\%$,
and ARGUS,~\cite{ARGUS_dilepton} $\slbr=(9.7\pm0.5\pm0.4)\%$, I obtain 
\begin{eqnarray}
\slbr=(10.23\pm0.39)\%.
\end{eqnarray}

For comparison, I also compute the average value of $B_{\rm SL}^Z$, 
which is the semileptonic branching fraction for the $b$-hadron
mixture produced in $Z$ decays.
I use measurements from 
ALEPH~\cite{ALEPH_BSL} [$(11.01\pm0.23\pm0.28\pm0.11)\%$], 
DELPHI~\cite{DELPHI_BSL} [$(11.06\pm0.39\pm0.19\pm0.12)\%$],
L3~\cite{L3_BSL} [$(10.85\pm0.12\pm0.43)\%$], 
and OPAL~\cite{OPAL_BSL} [$(10.5\pm0.6\pm0.4\pm0.3)\%$],
where the L3 value is a new result submitted to this conference, and
the ALEPH result has the lesser model dependence of the two 
measurements presented in their paper. I assume a common
systematic error of 0.25\% (absolute). The resulting average, 
$B_{\rm SL}^Z=(10.95\pm0.13\pm0.29)\%,$ is somewhat lower
than previous values of $B_{\rm SL}^Z$, and it is more consistent
with the value of $\slbr$ measured at the $\Upsilon(4S)$. 

To obtain averages of exclusive branching fractions from different
experiments, it is important to use a common
set of values for $D$ and $D^*$ branching fractions and,
in the case of the LEP experiments, for 
$R_b=\Gamma(Z\to b\bar b)/\Gamma(Z\to {\rm hadrons})$ 
and $f_{b\to B}$, the
fraction of $b$ quarks that hadronize into a $B^-$ 
or $B^0$ meson.~\cite{ALEPH_btodstlnu}
Table~\ref{tab:assumptions} lists the values that I have assumed for these
quantities. The $D^0\to K^-\pi^+$  branching
fraction is the average that I calculate below, while the
$D^+\to K^-\pi^+\pi^+$ branching fraction has been scaled to take 
into account its experimental dependence on the $D^0\to K^-\pi^+$ value.
The $D^{*+}\to D^0\pi^+$ branching fraction is taken from the 1994
Particle Data Book,~\cite{PDG94} 
but I have corrected the error, which was listed
incorrectly in that reference.

Table~\ref{tab:sum_modes} summarizes the status of
semileptonic branching fractions.
For $\btodstlnu$, I compute a world average based on measurements
at the $\Upsilon(4S)$ (CLEO~\cite{CLEO_rhosq} 
and ARGUS~\cite{ARGUS_rhosq,ARGUS_partial_rec}) and at the $Z$ 
(ALEPH~\cite{ALEPH_btodstlnu}, DELPHI~\cite{DELPHI_rhosq},
OPAL~\cite{OPAL_rhosq}).
For the semileptonic branching
fraction to $\btoclnu$ modes other than $\btodlnu$ and $\btodstlnu,$
I use a measurement from ALEPH based on a topological 
vertex study.~\cite{ALEPH_btodststlnu} This measurement determines
$B(b\to \bar B)\times B(\bar B\to D^{*+}\pi^-\ell^-\bar\nu X)=
(4.73\pm0.77\pm0.55)\times 10^{-3}$, as well as analogous 
branching fractions for final states with $D^{*0}\pi^+$, $D^+\pi^-$, 
and $D^0\pi^+$. Combining all of these results, ALEPH obtains
$B(\bar B\to D\pi X\ell^-\bar\nu+
\bar B\to D^*\pi X\ell^-\bar\nu),$ where both resonant and nonresonant
hadronic final states are included and $\btodlnu$ and $\btodstlnu$ are 
excluded. (I have rescaled their result using the new $D^0\to K^-\pi^+$
branching fraction.)
For $b\to u\ell^-\bar\nu$, I have made a rough estimate 
(with 50\% uncertainty) based on the value of $|V_{ub}|$ and the free
quark model prediction.~\cite{Rosner_btoulnu}
The sum of all these branching
fractions is $(9.38\pm0.65)\%,$  about $1.2\sigma$
lower than the inclusive semileptonic branching fraction.
This comparison shows that
a fairly large fraction of the semileptonic rate is accounted for. 
It is clear, however, the much work remains in improving the
precision of the branching fraction measurements and, in particular,
obtaining a detailed understanding of the $\btodststlnu$ modes
and modes with non-resonant final states.
\begin{table}[tb]
\begin{center}
\caption{Assumed values for quantites used to calculate averages
of semileptonic branching fractions.
\label{tab:assumptions}
\hspace*{\fill}}
\vspace{0.4cm}
{
\begin{tabular}{|cc|}\hline
$B(D^0\to K^-\pi^+)$      & $(3.88\pm0.10)\%$ \\
$B(D^+\to K^-\pi^+\pi^+)$ & $(8.8\pm0.6)\%$ \\
$B(D^{*+}\to D^0\pi^+)$      & $(68.1\pm1.6)\%$ \\
$f_{b\to B}$ (LEP)        & $(37.8\pm2.2)\%$ \\
$R_b$ (LEP)               & $(22.09\pm0.21)\%$ \\
\hline    
\end{tabular}
}
\end{center}
\end{table}
\begin{table}[tb]
\begin{center}
\caption{Contributions to the $B$ meson inclusive semileptonic
branching fraction.
\label{tab:sum_modes}
\hspace*{\fill}}
\vspace{0.4cm}
{
\begin{tabular}{|cc|}\hline
Mode                               & $B_i(\%)$         \\ \hline
$\bztodlnu$                        & $2.03\pm0.28$    \\
$\bztodstlnu$                      & $4.86\pm0.29$    \\
$[(\bar B\to D\pi X\ell^-\bar\nu)$    & $2.34\pm0.45$    \\
$+(\bar B\to D^*\pi X\ell^-\bar\nu)]$ &                  \\
$\bar B\to X_u\ell^-\bar\nu$       & $0.15\pm0.075$   \\ \hline
$\Sigma B_i$                       & $9.38\pm0.65$    \\
$\slbr(4S)$                        & $10.23\pm0.39$   \\
$\slbr(4S)-\Sigma B_i$             & $0.85\pm0.76$    \\ \hline 
\end{tabular}
}
\end{center}
\end{table}

The semileptonic branching fraction is also of great interest because
it tests our understanding of the hadronic rate. Although there are
significant theoretical uncertainties due to quark masses and
the renormalization scale, important progress has been made. 
The semileptonic branching fraction could also
be sensitive to an anomalously large rate for $b\to s g$, which
has been suggested by some theorists~\cite{Kagan_talk}. 

Historically,
most predictions of $B_{\rm SL}$ have been larger than the
values measured at the $\Upsilon(4S)$. Recent calculations performed
within the framework of $1/m_Q$ expansions have focused attention
on this issue. For example,
Bigi {\it et al.}`\cite{Bigi_slbr} have concluded that it is 
very difficult
to obtain a prediction for $\slbr$ below 12.5\%. Various
aspects of the problem have now been scrutinized in more detail, and
Bagan {\it et al.}~\cite{Bagan_nc1,Bagan_nc2} and Voloshin~\cite{Voloshin_nc}
have shown that higher order perturbative 
QCD corrections significantly increase the rate for $b\to c\bar c s$, 
thereby decreasing $\slbr$. 
To test this idea, one can measure $n_c$, the average number
of charm (or anti-charm) quarks per $B$ decay.
Thus, what was originally posed as the 
problem of the semileptonic branching fraction is now regarded
as the joint problem of $\slbr$ and $n_c$.
Bagan {\it et al.}~\cite{Bagan_nc2} obtain 
$\slbr=(12.0\pm0.7\pm0.5\pm0.2^{+0.9}_{-1.2})\%,$ where the
errors are due to uncertainties in $m_b$, $\alpha_s$, the $b$-quark
kinetic energy parameter $\lambda_1$, and the renormalization scale.
In an alternative scheme with ${\overline{\rm MS}}$ quark masses, they obtain  
$\bar B_{\rm SL}=(11.3\pm0.6\pm0.7\pm0.2^{+0.9}_{-1.7})\%$. They also
predict $n_c=1.24\pm0.05\pm0.01$, where the first error is due to 
the uncertainty in $m_b$. The alternative ${\overline{\rm MS}}$ calculation
gives $n_c=1.30\pm0.03\pm0.03\pm0.01$. 

\begin{figure}[tbh]
\epsfxsize=2.9in
\hbox{\hfill\hskip+0.1in
\epsffile{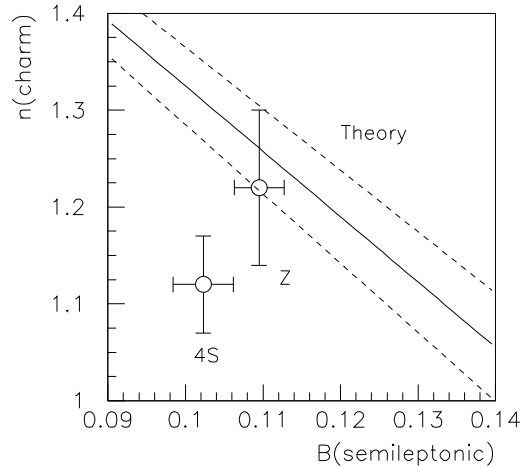}\hfill}
\vskip -3mm
\caption{Comparison of $n_c$ vs. $\slbr$ for measurements at 
the $\Upsilon(4S)$ and the $Z$ with theoretical prediction.
\hspace*{\fill}}
\label{fig:ncvsbsl}
\end{figure}

Experimentally, the quantity $n_c$ is difficult 
to determine because one must 
identify and measure all final states in $B$ decay that contain
one or more charm quarks. Inclusive
production of $D^0$, $D^+$, $D_s$, and $J/\psi$ is relatively
easy to measure, but the contributions from other charmonium
states and from charm baryons are more difficult to determine. 
To compare results from different experiments, it is important to
use a common set of charm branching fractions. I have updated the relevant
CLEO II and ARGUS $B$ branching fractions, which are 
listed in Browder {\it et al.}~\cite{Browder_rev}, to reflect the charm branching
fractions used in this article. In summing the $B$ branching fractions, I
take into account the correlated systematic error arising from common
charm branching fractions. I have applied the same procedure to 
new data from the ALEPH experiment.~\cite{ALEPH_nc} The values are
\begin{eqnarray}
n_c(4S)=1.12\pm0.05\qquad n_c^Z=1.22\pm0.08,
\end{eqnarray}
where $n_c^Z$ refers to the ALEPH measurement and 
is a reminder that the $b$ hadron content
at the $Z$ is more complicated than at the $\Upsilon(4S)$.
(This is seen, for example, in a significantly higher $D_s^+$
contribution at the $Z$ than at the $\Upsilon(4S)$.)
The error on the ALEPH measurement is larger than they report
because I use the newer $D_s^+\to\phi\pi^+$ branching 
fraction discussed later in this paper.
OPAL~\cite{OPAL_nc} has reported the value $n_c^Z=1.100\pm0.045\pm0.060,$ 
which I have not updated to the new $D^0$ and $D_s^+$ branching fractions.

Figure~\ref{fig:ncvsbsl} compares the points ($\slbr$,$n_c(4S)$) and
($\slbr^Z$, $n_c^Z$) with the theoretical summary given in
Buchalla {\it et al.}~\cite{Buchalla_nc}.
My view is that, given the size of the present experimental 
and theoretical uncertainties,
we cannot state that there is or is not a problem
in explaining the measured value of $\slbr$. 
There have already been substantial refinements in both experimental
and theoretical analyses of this question, and it is important
to continue these studies until the issue is resolved.

\section{Hadronic Decays of Charm and Bottom Mesons}
\subsection{Normalization modes for the $D^0$ and $D_s^+$}
\begin{figure}[tbh]
\epsfxsize=2.9in
\hbox{\hfill\hskip+0.1in
\epsffile{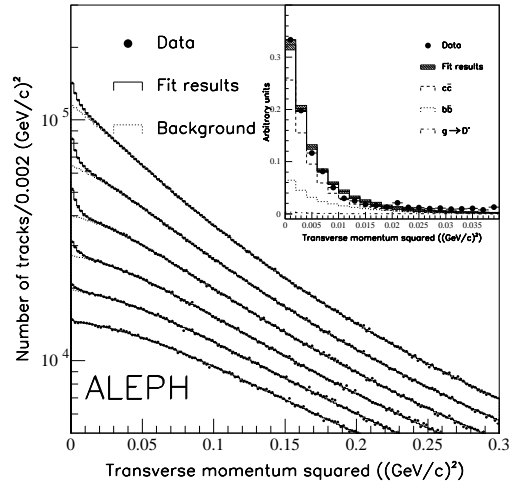}\hfill}
\vskip -5mm
\caption{ALEPH measurement of $B(\dtokpi)$. The data points show
the distribution of transverse momentum squared of hadrons,
relative to the axis of the nearest jet, in
six pion momentum bins from 1 GeV/$c$ at the top to 4 GeV/$c$ at the
bottom. The peaks at the low end of the distributions are 
due to the soft pion from the decay $D^{*+}\to D^0\pi^+.$
The figure in the upper-right corner shows the 
background-subtracted $p_T^2$ distribution after
the data from all six pion momentum bins are combined.
\hspace*{\fill}}
\label{fig:ALEPH_dtokpi}
\end{figure}

The branching fractions for the decays $\dtokpi$ and $\dstophipi$ 
are crucial for many measurements in charm and bottom physics,
since they ultimately determine the absolute branching fraction
scales. A new high-precision measurement of $\dtokpi$ from ALEPH
was submitted to this conference.~\cite{ALEPH_dtokpi} The key
to this analysis, like those of earlier measurements from
the HRS,~\cite{HRS_dtokpi} ARGUS,~\cite{ARGUS_dtokpi} and 
CLEO,~\cite{CLEO_dtokpi} is to tag $D^{*+}\to D^0\pi^+$
without using the $D^0$ decay products. The tag is based on
the fact that the energy release in the $D^{*+}$ decay is
extremely low, about 6 MeV. In the $D^{*+}$ rest frame, the
resulting ``soft pion'' has momentum 
$p_{\pi}^*\approx 40\ {\rm MeV}/c,$ which limits
its transverse momentum relative to the $D^{*+}$ 
direction in the lab frame. This low $p_T$ pion
provides a distinctive signature for the $D^{*+}$
decay. Since the $p_T$ measurement must be performed
without reconstructing the full $D^{*+}$ decay chain, the
$D^{*+}$ direction is approximated by the jet axis.
To determine $B(D^0\to K^-\pi^+)$, one then
measures the fraction of such tags in which the
decay $D^0\to K^-\pi^+$ is found.
A key part of this measurement is to
accurately determine the background shape in $p_T$ that arises 
from soft pions due to other sources. Figure~\ref{fig:ALEPH_dtokpi}
shows the ALEPH $p_T^2$ distributions, which have peaks at the
low end due to $D^{*+}\to D^0\pi^+$.
\begin{figure}[tbh]
\epsfxsize=3.0in
\hbox{\hfill
\epsffile{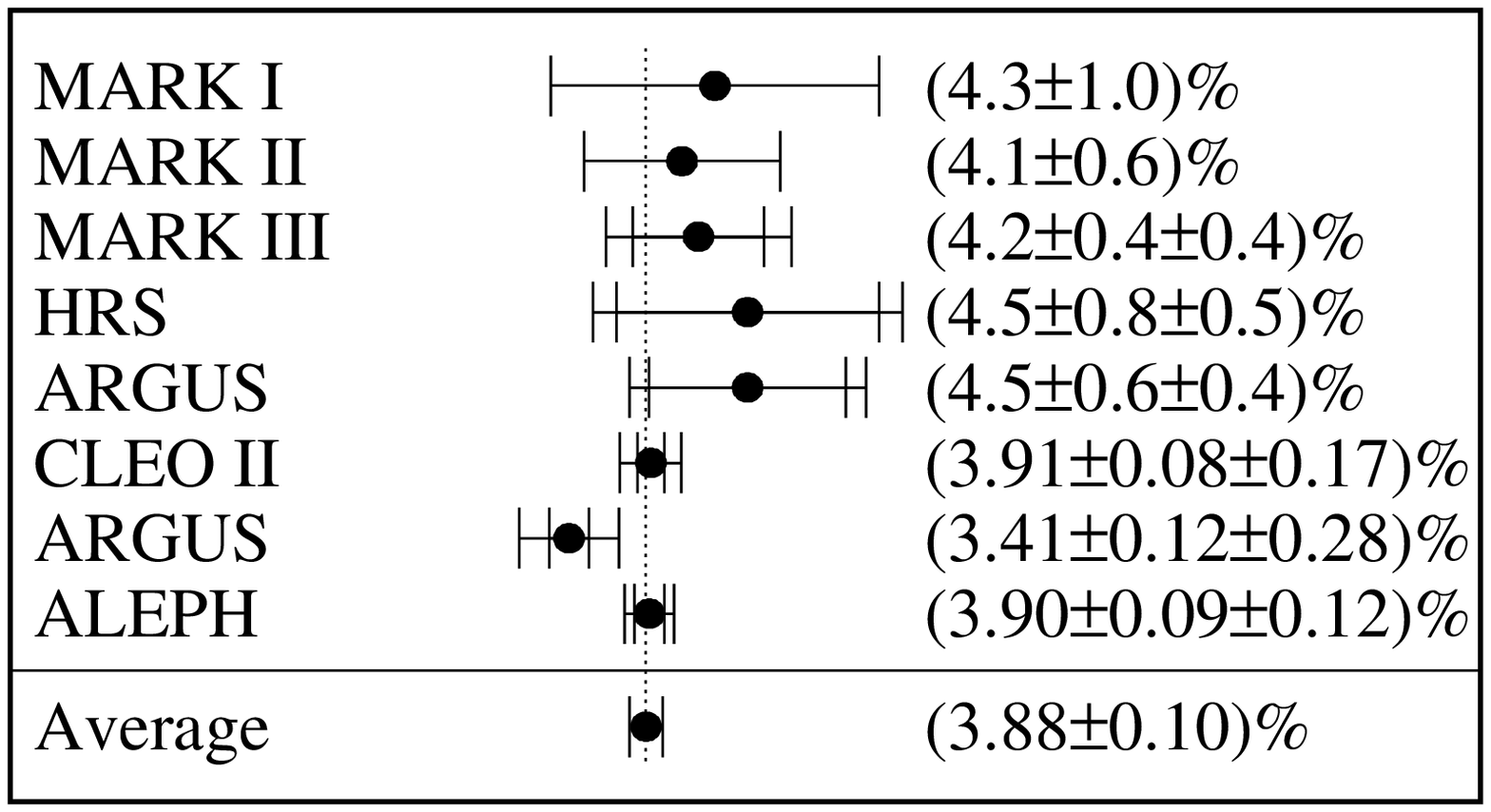}\hfill}
\vskip -3mm
\caption{Summary of measurements of $B(\dtokpi).$\hspace*{\fill}}
\label{fig:summary_dtokpi}
\end{figure}

Figure~\ref{fig:summary_dtokpi}
summarizes the status of $\dtokpi$ measurements. 
The new preliminary result from ALEPH is somewhat more precise
than the CLEO II measurement, and the two values are quite
consistent. The average of all the measurements is 
$B(\dtokpi)=(3.88\pm0.10)\%,$ which I use in several
places in this paper.   
Recently, Dunietz~\cite{Dunietz_dtokpi} has argued that
a lower value of this branching fraction is indicated by a number
of problems in $B$ physics. This conclusion is not supported by 
the new ALEPH result, but it is extremely important that
the $\dtokpi$ branching fraction be checked by other experiments.

Measurements of $B(\dstophipi)$ are much less precise.
The 1994 version of the Particle Data Book~\cite{PDG94} lists 
$B(\dstophipi)=(3.5\pm0.4)\%$. However, this value is 
based mainly on measurements of
$\Gamma(D_s^+\to\phi\pi^+)/\Gamma(D_s^+\to\phi\ell^+\nu).$
The scale is then set by assuming that    
$\Gamma(D_s^+\to\phi\ell^+\nu)$ and $\Gamma(D\to K^*\ell^+\nu)$
are equal, up to a small theoretical correction.
The new CLEO II measurement~\cite{CLEO_dstophipi}
of the absolute branching fraction 
avoids any theoretical assumptions.
In this measurement, the decay $\bar B^0\to D^{*+}D^{*-}_s$
is reconstructed in two ways: (1) the $D_s^{*-}$ is fully
reconstructed and the $D^{*+}$ is partially reconstructed using
the soft pion from $D^{*+}\to D^0\pi^+$, and (2) the $D^{*+}$
is fully reconstructed and the $D_s^{*-}$ is partially
reconstructed using the soft photon from 
$D_s^{*-}\to D_s^-\gamma$.
CLEO obtains $B(\dstophipi)/B(\dtokpi)=0.92\pm0.20\pm0.11,$ which is the
basis for the 1996 Particle Data Book~\cite{PDG96} 
value $B(\dstophipi)=(3.6\pm0.9)\%.$
The quoted uncertainty is significantly larger than before, 
but no theoretical assumptions are made. 
\subsection{Hadronic $B$ Decays and Factorization}

Hadronic decays are much more complicated than leptonic
or semileptonic modes, because all of the fermions involved
are quarks and can interact strongly.  
Although it is not possible to make precise predictions for hadronic decays,
the factorization hypothesis provides a framework for
understanding many of the observed features of two-body modes.
In the factorization approach, one writes the decay amplitude as
the product of two currents, in analogy to semileptonic decay.
Factorization was discussed extensively at this conference by several 
experimentalists,~\cite{Browder_talk,CDFtalk,Skwarnicki_talk} and these
talks led to considerable discussion regarding the applicability
and reliability of factorization in various decay processes.
Here, I will only give an introduction to some
of the experimental results related to this complex topic. 

\begin{figure}[tbh]
\epsfxsize=2.9in
\hbox{\hfill\hskip+0.1in
\epsffile{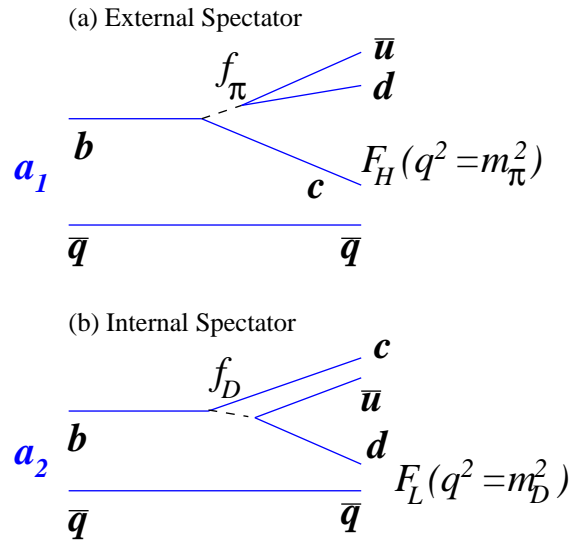}\hfill}
\vskip 2mm
\caption{Spectator diagrams for $B$ decays with 
(a) an external $W$ and (b) an internal $W$. The QCD
hard-gluon corrections are parametrized by the
coefficients $a_1$ and $a_2$, while nonperturbative
effects are described, in our $B\to D\pi$ example,
by the decay constant $f_{\pi}$
or $f_D$ and the form factor $F_H(q^2=m_{\pi}^2)$ 
or $F_L(q^2=m_{D}^2).$ \hspace*{\fill}}
\label{fig:factorization}
\end{figure}

Figure~\ref{fig:factorization} shows the hadronic decay of a $B$
meson through (a) an external spectator diagram and (b) an internal
(color suppressed) spectator diagram. 
In the factorization approach, the effects of strong interactions
are divided into two categories: (1) short-distance, hard-gluon effects
parametrized by the coefficient $a_1$ for the external diagram and 
$a_2$ for the internal diagram and (2) long-distance, soft-gluon effects,
which are parametrized by decay constants and form factors.
In Fig.~\ref{fig:factorization}(a) the $\bar u d$ system
from the $W$ decay is produced at a point, so that the appropriate meson
decay constant, which measures the overlap of the quark anti-quark pair,
parametrizes the amplitude to produce the meson.
To be definite, we take this meson to be a pion.
The daughter charm quark 
recoiling at the lower $W$ vertex, however, must bind together with the 
spectator quark. The physics of this process is very similar to the
hadronic transition in semileptonic decay discussed earlier,
and it is described
by the appropriate form factor evaluated at $q^2=m_{\pi}^2$. Thus, for
the external spectator diagram example, the non-perturbative QCD physics
is described by the pion decay constant $f_{\pi}$ and
a ``heavy-to-heavy'' ($B\to X_c$) form factor $F_H(q^2=m_{\pi}^2)$. 

For $a_1$-type decays with large daughter-hadron recoil 
velocity (i.e., decays at low $q^2$), 
one might expect factorization to be a good approximation from the
following argument, due to Bjorken.~\cite{BJ_factorize} 
The $q\bar q$ pair produced in
the $W$ decay is formed at a point as a color singlet, and, for low $q^2$
processes, the pair moves out of the decaying hadron at high velocity
in a collinear fashion. 
Thus, the pair looks like a very
small color dipole that gradually grows to the size of a meson.
The pair will not form a meson, however, 
until it moves a distance $\approx\gamma c\tau_h$,
where $\tau_h$ is a typical hadronization time in the rest frame,
$\tau_h\sim 1\ {\rm fm}/c$. This distance can be as large as 20 fm,
much larger than the decaying meson, 
so that the pair can escape from the cloud of quarks
and gluons without significantly
interacting with it.

Consider now the internal spectator diagram, shown in
Fig.~\ref{fig:factorization}. Because the $W$ has such a short range
compared with the size of a hadron, the $c$ and the $\bar u$ are still
effectively produced at the same point, and the relevant decay constant
is $f_D$ instead of $f_{\pi}$. It is the daughter
$d$ quark that now must bind with the
spectator, so there is now a ``heavy-to-light'' ($B\to X_u$) 
form factor, which I denote
by $F_L(q^2=m_D^2).$

For a $\bar B^0$ decay,
the external and internal diagrams lead to different final states
(charged+charged and neutral+neutral), whereas in $B^-$ decay two
$\bar u$ quarks are present in the final state, which means that the
same charged+neutral final state can be reached in two ways. Thus,
we have the pattern
\begin{eqnarray}
\bar B^0(b\bar d)&\to& {\rm chg}\ + {\rm chg}:\ 
         \Gamma\propto |a_1 f_{\pi} F_H |^2  \nonumber\\ 
\bar B^0(b\bar d)&\to& {\rm neut}\ + {\rm neut}:\ 
         \Gamma\propto |a_2 f_{D} F_L |^2  \nonumber \\ 
     B^-(b\bar u)&\to& {\rm chg}\ + {\rm neut}:\nonumber \\  
      & & \Gamma\propto |a_1 f_{\pi} F_H+ a_2 f_D F_L|^2  
\end{eqnarray}

Using measured branching fractions for hadronic $B$ decays, one can
extract the coefficients $a_1$ and $a_2$ arising from hard gluon
effects. This procedure requires knowledge of both decay
constants and form factors. As discussed earlier, our
knowledge of these quantities is far from perfect, especially 
in the case of the form factors for heavy-to-light 
transitions. Compared with the state of our understanding of hadronic
decays, however, form factor predictions might be regarded as reasonably
trustworthy, but it is important to remember that there
is more uncertainty in calculations made within the factorization
framework than is often acknowledged.

One approach to testing factorization is to compare the decay rate
for a hadronic mode with the rate at the same value of $q^2$
for a semileptonic decay. 
For $B\to D^{(*)}P$ decays, where
$P$ is a pseudoscalar meson, one can define~\cite{Neubert_factorize} 
the ratio
\begin{eqnarray}
R_P^{(*)}&=&{\Gamma(\bar B\to D^{(*)}P)\over
d\Gamma(\bar B\to D^{(*)}\ell^-\bar\nu)/dq^2|_{q^2=m_P^2}}\nonumber \\ 
&=& 6\pi^2 f^2_P\, a_1^2 |V_{ij}|^2X_P^{(*)}(q^2=m_P^2),
\label{eq:factorization}
\end{eqnarray}
where $X_P^{(*)}$ is the ratio of form factors 
for the hadronic to the semileptonic decay.  
(This ratio is defined in 
Neubert {\it et al.}\,\cite{Neubert_factorize} In most cases,
$X_P^{(*)}$ is approximately equal to one.
However, the semileptonic decay form factor 
that becomes important only for large lepton mass is the one that
enters into the hadronic decay, so there can be some subtleties here.) 

The upper line of Eq.~\ref{eq:factorization} can be evaluated
from experiment, while the lower line can be calculated from theory,
the comparison giving a test of whether factorization is valid.
The constant $a_1$ can be calculated from QCD: $a_1=c_1+c_2/3\approx 1.0$,
where $c_1$ and $c_2$ are calculated using the renormalization
group equation. However, I prefer to use Eq.~\ref{eq:factorization} to
extract $a_1$ for different decay modes and then to check whether
the resulting values are consistent.

\begin{table}[tb]
\begin{center}
\caption{Values of $a_1$ extracted from a comparison of hadronic and
semileptonic $B$ decays. The consistency of these values is an
indication of whether factorization provides a good description of the
hadronic decays.
\hspace*{\fill}}
\label{tab:factorization}
\vspace{0.4cm}
\begin{tabular}{|lc|}\hline
Mode       &  $a_1$  \\ \hline
$\bar B^0\to D^+\pi^-$      & $0.92\pm0.10$ \\
$\bar B^0\to D^+\rho^-$     & $0.97\pm0.13$ \\ \hline
$\bar B^0\to D^{*+}\pi^-$   & $0.96\pm0.10$ \\
$\bar B^0\to D^{*+}\rho^-$  & $0.92\pm0.12$ \\
$\bar B^0\to D^{*+}a_1^-$   & $1.1\pm0.13$ \\ \hline
$\bar B^0\to D^{(*)+}D_s^{(*)-}$& $1.24\pm0.40$\\
\hline    
\end{tabular}
\end{center}
\end{table}

Table~\ref{tab:factorization} shows the values of $a_1$ for
a set of $\bar B^0$ decay modes that can be compared with $\bztodlnu$
and $\bztodstlnu$. The agreement among these
values is reasonably good, indicating that at the low values
of $q^2$ (fast recoil) at which these hadronic decays occur,
factorization provides a good description of the process.
The data for the comparison with $\btodlnu$ are taken from the
CLEO~\cite{CLEO_btodlnu}, while the $\btodstlnu$ comparison is made
using combined CLEO and ARGUS data compiled in 
Browder {\it et al.}~\cite{Browder_rev}. The data for the
$\bar B^0\to D^{(*)+}D_s^{(*)-}$ are from CLEO~\cite{CLEO_BtoDsX}
and assume $f_{D_s}=f_{D_s^*}.$ 

Another interesting test of factorization is to compare the
polarization of the $D^*$ in the hadronic decay 
$\bar B^0\to D^{*+}\rho^-$ with that
in the semileptonic decay $\bztodstlnu$, when measured at the same value of
$q^2$. In a semileptonic
decay at low $q^2,$ the lepton and antineutrino are nearly collinear, 
so their net spin along their direction of motion is zero. Since
the $B$ has spin zero, the recoiling meson must also have
helicity zero. In the factorization picture, the quark-antiquark
pair from the $W$ would behave like the lepton-antineutrino system,
so that the $\rho^-$ (or $D^{*+}$) is expected to
be almost completely longitudinally polarized. 
The new CLEO II measurement~\cite{CLEO_btodstlnuff_pre} 
of the $\btodstlnu$ form factors gives
$(\Gamma_L/\Gamma)_{q^2=m_{\rho}^2}=(91\pm1.3\pm0.6)\%$.
For $\bar B^0\to D^{*+}\rho^-$,
CLEO obtains~\cite{CLEO_bigB} $\Gamma_L/\Gamma=(93\pm5\pm5)\%,$
a very similar value.

We can also compare the decay rates for $\bar B^0\to D^{*+}\rho^-$
and $\bar B^0\to D^+\rho^-$. One might expect the rate for the 
to $D^{*+}\rho^-$ to be larger, because there are more spin states
accessible. However, factorization predicts that the rates should
be about the same, since the $D^{*+}$ polarization is almost
completely longitudinal (helicity zero). 
The PDG96~\cite{PDG96} values are
$B(\bar B^0\to D^{*+}\rho^-)=(0.73\pm0.15)\%$ and
$B(\bar B^0\to D^{+}\rho^-)=(0.78\pm0.14)\%$, which are indeed very similar.

Although factorization is on a less secure footing for 
color-suppressed processes, 
there is great interest in the ratio $a_2/a_1$, whose sign
manifests itself in the interference term in $B^-$ decays.
The magnitude of $a_2$ can also be determined from 
color-suppressed $\bar B^0$ decays, such as $\bar B^0\to J/\psi K^{0(*)}$.
Skwarnicki~\cite{Skwarnicki_talk} presented new CLEO II
measurements of $B(\bar B^0\to D^{*+}\pi^-)$ and
$B(B^-\to D^{*0}\pi^-)$ that are based on a novel partial
reconstruction technique, resulting in smaller statistical errors.
From the ratio of these branching fractions, he 
extracts $a_2/a_1=+(0.27\pm0.03)$, which differs substantially
from the value in
charm decays, where $a_2/a_1$ is negative. As discussed earlier,
the error does not include all theoretical uncertainties. 
The long $D^+$ lifetime (relative to that of the $D^0$ or $D_s^+$)
is attributed to the negative value of $a_2/a_1;$  
the positive value in $B$ decays would indicate
a shorter $B^+$ than $B^0$ lifetime, although the importance of
two-body decays (from which this ratio is obtained) may be
less in $B$ decays.

\subsection{Rare Hadronic Decays}
Experiments are now achieving sensitivities to $B$ branching fractions
in the range $10^{-4}$ to $10^{-5}$, opening up new
types of processes for study. Here I will focus on rare decays
to hadronic final states; electromagnetic penguins are covered in
the talk by A.~Buras. 

Many rare hadronic decays, such as $B\to\pi\pi$ and $B\to K\pi$, can 
proceed either through a $b\to u$ spectator diagram or through a
gluonic penguin. In the latter process, a $b\to d$ or $b\to s$ 
transition occurs
through a virtual loop containing a $W$ and either a $t$ or $c$ quark,
with the radiation of a gluon. The theoretical expectation is
that the decays $\bar B^0\to\pi^+\pi^-$ and $\bar B^0\to \pi^+\rho^-$ 
are dominated by the $b\to u$ spectator process, while 
$\bar B^0\to K^-\pi^+$ and $\bar B^0\to K^{*-}\pi^+$ are dominated by
$b\to s g$. The gluonic penguins are of interest not only as one-loop
processes, but also because they can affect studies of
CP violation.
For example, the final state in 
$\bar B^0(B^0)\to \pi^+\pi^-$ is an eigenstate of
the CP operator, and this decay is well suited to the measurement of
$\sin 2\alpha$, where $\alpha$ is one of the angles of the 
CKM unitarity triangle. However, a significant contribution 
to this mode from $b\to d g$ 
would complicate the interpretation of the CP asymmetry and is
sometimes called ``penguin pollution.'' 
Measurements of the relative size of branching fractions
to non-strange and strange final states would provide information
on the level of this contamination and are therefore of great 
interest for CP violation studies.

A simple argument gives some idea of the possible
level of penguin contamination.
Assuming that the $t$-quark dominates the loop, the 
penguin contribution to $\bar B^0\to\pi^+\pi^-$ is
suppressed relative to that for $\bar B^0\to K^-\pi^+$ by
$|V_{td}/V_{ts}|^2={\cal O}(\lambda^2)$, where
$\lambda\simeq 0.22$ is the sine of the Cabibbo angle.
Furthermore, the spectator contribution to $\bar B^0\to K^-\pi^+$ is
suppressed relative to that for $\bar B^0\to\pi^+\pi^-$
by a factor $|V_{us}/V_{ud}|^2\simeq \lambda^2$.
Now, suppose that
$B(\bar B^0\to\pi^+\pi^-)\approx B(\bar B^0\to K^-\pi^+)$,
as is crudely indicated by experiment. Then our CKM
argument implies that $\bar B^0\to K^-\pi^+$ must be
mainly penguin, or else the $\pi^+\pi^-/K^-\pi^+$ ratio
would be larger. Furthermore, even if all of the
$\bar B^0\to K^-\pi^+$ rate were penguin, the (assumed)
near equality of the branching fractions implies that
the penguin contribution to $\bar B^0\to\pi^+\pi^-$ must be
fairly small, since this contribution is suppressed by
$\lambda^2$ relative to the penguin contribution to  
$\bar B^0\to K^-\pi^+$.
%
\begin{figure}[tbh]
\epsfxsize=2.9in
\hbox{\hfill
\epsffile{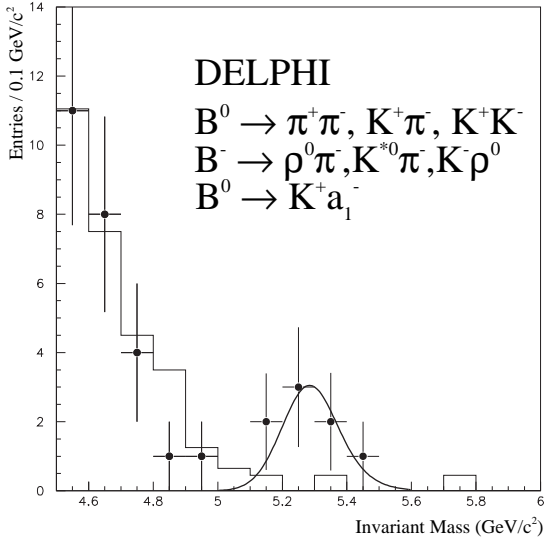}\hfill}
\vskip -5mm
\caption{DELPHI data on two-body hadronic $B$ decays to charmless
final states. The points with error bars show the distribution of
invariant masses for the data; the histogram shows the expectation
if there were no charmless $B$ decays, as predicted from Monte Carlo.
The curve represents the signal shape normalized to the data.\hspace*{\fill}}
\label{fig:DELPHI_rare}
\end{figure}

CLEO has updated~\cite{CLEO_bigrare} its original measurement of 
the sum of rates to $\pi^-\pi^+$ and $K^-\pi^+$:
\begin{eqnarray}
B(\bar B^0&\to &\pi^-\pi^++K^-\pi^+) \nonumber\\ 
&=&(1.8^{+0.6+0.2}_{-0.5-0.3}\pm0.2)
\times 10^{-5}.
\end{eqnarray}
There are about 17 events in the signal, with a significance of 
$4\sigma$ to $5\sigma$. Due to the high momentum of the decay products,
the $\pi/K$ separation is difficult, and
the two modes are not clearly distinguished. CLEO reports
$R=N_{\pi\pi}/(N_{\pi\pi}+N_{K\pi})= \qquad\qquad\qquad
0.54^{+0.19}_{-0.20}\pm0.05$. The systematic error on $R$
is relatively small, so it will be possible to substantially
improve the separation between the modes with 
additional data.

\begin{figure}[tbh]
\epsfxsize=2.9in
\hbox{\hfill
\epsffile{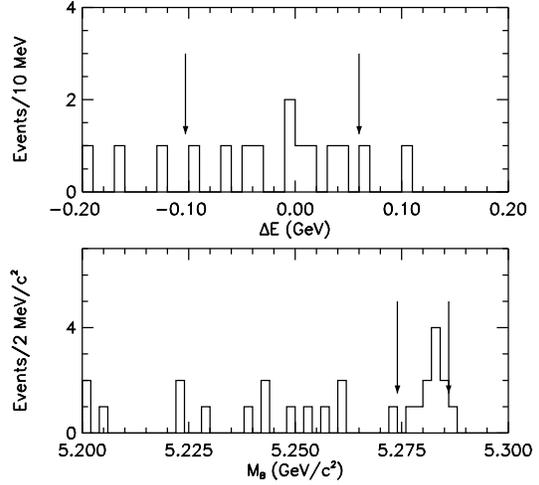}\hfill}
\vskip -2mm
\caption{CLEO II measurement of $B^-\to\omega h^{-}$.
The upper histogram shows the distribution of $\Delta E$,
the difference between the total energy of the candidate
particles and the beam energy. The lower histogram shows
the distribution of beam-energy constrained masses. The arrows
indicate the signal regions in these two variables.\hspace*{\fill}}
\label{fig:CLEO_omegah}
\end{figure}
%
DELPHI and ALEPH both presented new results on rare  
$b$ hadron decays. The DELPHI analysis~\cite{DELPHI_rare} benefits from good
particle identification from their ring-imaging Cerenkov detector
(RICH). They observe eight events, of which five are in 
the two-prong subsample ($\pi^+\pi^- + K^+\pi^-$) and three are in the
three-prong subsample ($\rho\pi+K^*\pi$). Within the two-prong subsample,
three events are from $K^+\pi^-$. DELPHI also determines that the
$B_s$ contribution to this sample is relatively small, about 1.3 events.
They obtain 
$B(B^0_{d,s}\to\pi^+\pi^-, K^+\pi^-)=(2.8^{+1.5}_{-1.0}\pm0.2)\times
10^{-5}$, consistent with CLEO, and 
\begin{eqnarray}
{(B_{u,d}\to K\pi, K^*\pi)\over (B_{u,d}\to\pi\pi,\rho\pi)+
(B_{u,d}\to K\pi, K^*\pi)} \nonumber\\  
\qquad =0.58\pm0.18.
\end{eqnarray}
Figure~\ref{fig:DELPHI_rare} shows the invariant mass spectrum from DELPHI
for candidate two- and three-prong events.
ALEPH~\cite{ALEPH_rare} also presented results 
from a search for rare hadronic decays;
they find four signal events and obtain
$B(B\to h^+ h^-)=(1.7^{+1.0}_{-0.7}\pm0.2)\times 10^{-5}$,
where the initial state can include $B^0$, $B_s$, and $\Lambda_b$.

CLEO presented evidence~\cite{CLEO_btoomegah} 
for a new rare decay, $B^+\to\omega h^+$, where
$h^+=\pi^+$ or $K^+$.  Figure~\ref{fig:CLEO_omegah} shows the
distributions of $\Delta E=E_{\omega}+E_h-E_{\rm beam}$ and
$M(\omega h^+)=\sqrt{E_{\rm beam}^2-({\bf p}_{\omega h})^2}$;
these variables have resolutions
$\sigma_{\Delta E}\approx 30\ {\rm MeV}$
and $\sigma_{M_B}\approx 3\ {\rm MeV}.$
A peak is evident in the $M(\omega h^+)$ distribution 
at the $B$ mass, as is a broad cluster of events centered
around $\Delta E=0$. 
The arrows in the plots indicate the signal region, which
contains a total of 10 events with an estimated background
(due to continuum) of 2 events. CLEO obtains the
preliminary result
$B(B^+\to \omega h^+)=(2.8\pm 1.0\pm0.6)\times 10^{-5}.$

Many other results were presented on rare $B$ decays, including an
update on $b\to s\gamma$ branching fractions from CLEO,~\cite{CLEO_btosgamma}
limits on $b\to s\gamma$ and $b\to s\nu\bar\nu$ 
from DELPHI,~\cite{DELPHI_rare} limits on inclusive modes
sensitive to $b\to s g$ from CLEO,~\cite{CLEO_btosgluon} and limits
on other hadronic rare decays from CLEO.~\cite{CLEO_btopp,CLEO_btohhh}
The study of rare $B$ decays is at an early stage, with only
handfuls of events observed so far in exclusive modes. As very large
data samples are accumulated at $B$ factories in the future,
we can expect many new results on these decays.

\section{Lifetimes of $b$ Hadrons}

Measurements of $\bar B^0$ and $B^-$ lifetimes  
are reaching an impressive precision, with  
systematic errors as small as $\pm0.02\ {\rm ps}$ (CDF), 
and total errors for individual measurements as
small as $\pm0.06\ {\rm ps}$ (DELPHI). 
There are also
substantial improvements in the lifetime
measurements for both the $B_s$ and $\Lambda_b$.

Predictions for the ratios of $b$-hadron lifetimes have been made
by many theorists. I will make only a few comments,
leaving the detailed theoretical discussion to the talk
by Martinelli.
Using the heavy-quark expansion,
Neubert obtains~\cite{Neubert_BandCP}
\begin{eqnarray}
{\tau(B^-)\over\tau(B^0)}&=&1+{\cal O}(1/m_b^3),\nonumber\\
{\tau(B_s)\over\tau(B^0)}&=&(1.00\pm0.01)+{\cal O}(1/m_b^3),\nonumber\\
{\tau(\Lambda_b)\over\tau(B^0)}&\simeq&0.98+{\cal O}(1/m_b^3),
\end{eqnarray}
where the estimate for $\tau(\Lambda_b)/\tau(B^0)$ includes
corrections that arise at order $1/m_b^2$.
Although these ratios might appear very close to unity,
Neubert argues that the $1/m_b^3$ corrections 
to $\tau(B^-)/\tau(B^0)$ 
might be large (due to phase-space enhancement of effects
involving the spectator quark), and he concludes that theoretical
uncertainties allow a lifetime ratio in the range
$0.8<\tau(B^-)/\tau(B^0)<1.2.$ This subject is
controversial, and other theorists have placed
tighter contraints on this ratio. For example, Bigi~\cite{Bigi_rev}
concludes that the $B^-$ lifetime is definitely longer than the
$B^0$ lifetime:
\begin{eqnarray}
{\tau(B^-)\over\tau(B^0)}\simeq 1+0.05\cdot{f_B^2\over (200\ {\rm MeV})^2}.
\end{eqnarray}
There is more consensus on $\tau(B_s)/\tau(B^0)$,
which is expected to be unity up to corrections of order 1\%.
Theoretical estimates for $\tau_{\Lambda_b}/\tau_{B^0}$
are typically in the range 0.9 to 1.0.

\begin{figure}[tbh]
\epsfxsize=2.9in
\hbox{\hfill
\epsffile{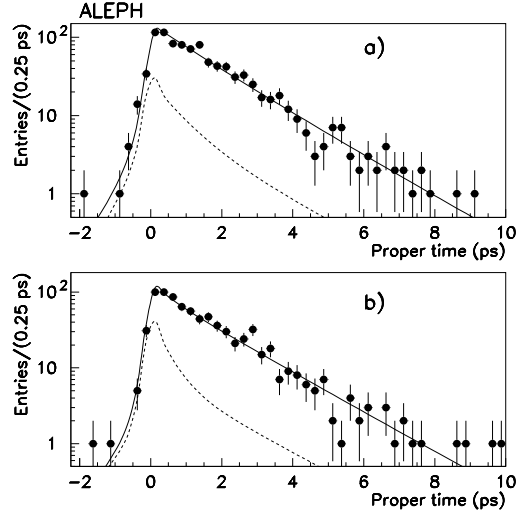}\hspace*{\fill}}
\vskip -4mm
\caption{Proper time distribution for the 
ALEPH measurement of $\tau(B^-)$ and $\tau(B^0)$ using
semileptonic $B$ decay. The points with errors show the data
for (a) $D^{*+}\ell^-$ events and (b) $D^0\ell^-$ events, 
which correspond mainly the $B^0$ and $B^-$ decays. The 
dashed curves represent the background contribution,
and the solid curves show the total fit.\hspace*{\fill}}
\label{fig:ALEPH_Blife}
\end{figure}

Lifetime measurements can be grouped into three broad categories,
corresponding to the use of semileptonic decays, fully reconstructed
hadronic decays, and inclusive methods such as topological vertexing. 
The main advantages of using semileptonic decays, such 
as $\bar B\to D^{(*)}\ell^-(X)\bar\nu,$ are the large
branching fractions, the presence of the lepton, 
and good vertex determination.
There are, however, significant disadvantages. Because
there is always at least one missing particle, the neutrino, 
it is generally not possible to reconstruct a $B$ mass peak. As a 
consequence, it can be difficult to determine whether
there are additional missing particles, especially neutrals.
The neutrino and other missing particles degrade the $B$ momentum
resolution and hence reduce the precision of the  
proper decay time measurement. In addition, missing charged particles
hinder the clean separation of $\bar B^0$ and $B^-$ samples.
Figure~\ref{fig:ALEPH_Blife}
shows the proper time distributions 
from ALEPH~\cite{ALEPH_Blife} for $D^{*+}\ell^-$ and
$D^0\ell^-$ samples, which would ideally correspond to $\bar B^0$
and $B^-$ decays. In reality, the $D^{*+}\ell^-$ sample is
$(87\pm 4)\%$ pure, and the $D^0\ell^-$ is $(75\pm5)\%$ pure,
so these distributions are fit simultaneously.
A model is need to account for possible feed-down contributions
from various $D^{**}\ell^-\bar\nu$ modes. In spite of these
complications, ALEPH achieves a very good systematic error
on these measurements:
$\tau_{B^0}=(1.61\pm0.07\pm0.04)\ {\rm ps}$ and
$\tau_{B^-}=(1.58\pm0.09\pm0.04)\ {\rm ps}$. 
DELPHI~\cite{DELPHI_Bzlife} reports
a precise preliminary result 
$\tau(B^0)=(1.529\pm0.040\pm0.041)\ {\rm ps}$
based on a study of $\bar B^0\to D^{*+}X\ell^-\bar\nu$ decays. Their 
measurement is based on a novel method in which the $D^0$ from
the $D^{*+}\to D^0\pi^+$ is inclusively reconstructed, leading
to a sample of $3520\pm150$ decays.

\begin{figure}[tbh]
\epsfxsize=2.9in
\hbox{\hfill\hskip 6mm
\epsffile{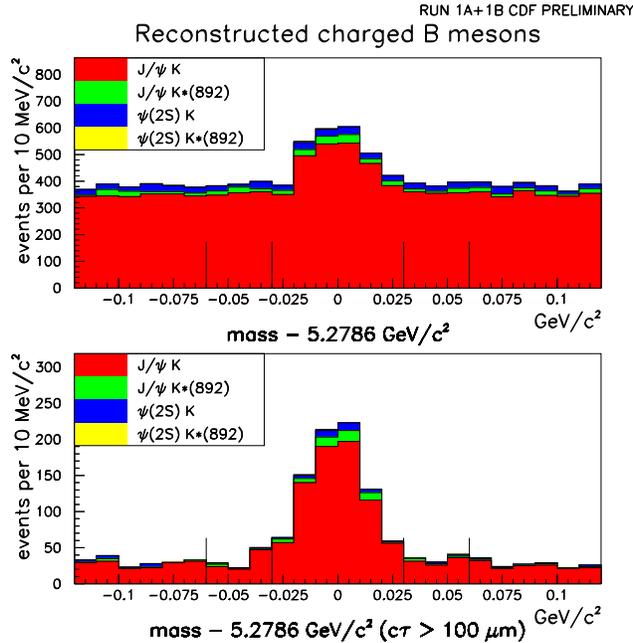}\hspace*{\fill}}
\vskip -12mm
\caption{CDF measurement of $\tau(B^+)$. The upper histogram (a)
shows the invariant mass distributions for $B^+$ candidates 
in the $J/\psi K^+$, $J/\psi K^{*+}$, $\psi(2S)K^+$ and 
$\psi(2S)K^{*+}$ final states; (b) shows the same distributions
after the cut $c\tau>100\ \mu{\rm m}$. 
\hspace*{\fill}}
\label{fig:CDF_mass_peaks}
\end{figure}
\begin{figure}[tbh]
\epsfxsize=2.9in
\hbox{\hfill\hskip 6mm
\epsffile{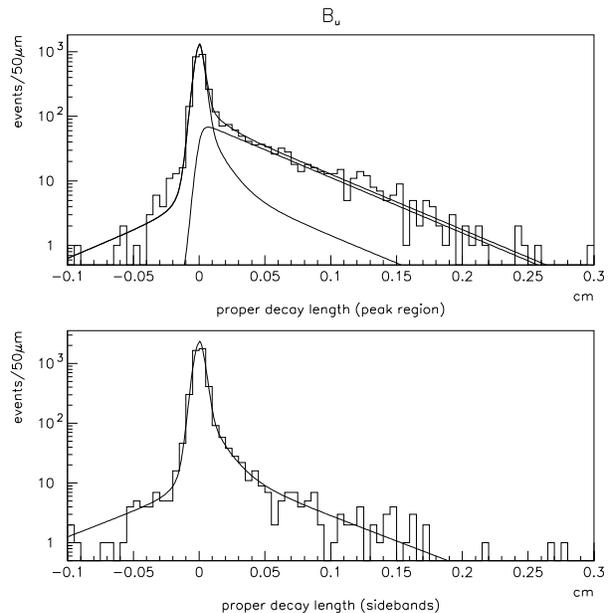}\hfill}
\vskip -12mm
\caption{CDF measurement of $\tau(B^+)$. The upper histogram
shows the proper decay length distribution for events in the $B$ mass peak;
the lower histogram shows the distribution for events in the
$B$ mass sidebands.\hspace*{\fill}}
\label{fig:CDF_lifetime_dist}
\end{figure}

The second method for determining $b$-hadron lifetimes uses 
fully reconstructed, exclusive hadronic decays, and the
analyses are usually 
much simpler than those using semileptonic modes. 
A $b$-hadron mass peak is observed, whose sidebands can be
used to study the lifetime distribtion of the background. Since
all particles are observed, the decay vertex and the 
$b$ hadron momentum are well determined, and 
the conversion to proper lifetime is very straightforward.
The only disadvantage
is that the event samples are typically smaller than those for
semileptonic decay. In hadron colliders, however, a sufficiently
large number of $b$ hadrons is produced for exclusive hadronic
decays to final states with $J/\psi$'s provide a competitive
method for measuring lifetimes. Figure~\ref{fig:CDF_mass_peaks} shows
the reconstructed $B^+$ mass from CDF~\cite{CDF_Blife}
for the final states $J/\psi K^+$,
$J/\psi K^{*+}$, $\psi(2S) K^+$, and $\psi(2S) K^{*+}$. The
upper and lower histogram in the figure show the data before and
after the proper decay length cut $c\tau>100\ {\mu}\rm m$.
This cut is not actually used for the lifetime fit, but it
demonstrates that the background is concentrated at short proper lifetimes.
Figure~\ref{fig:CDF_lifetime_dist} shows the distribution of 
proper decay lengths in the signal region (upper histogram) and in
the $B$ mass sidebands (lower histogram), together with the fits
to the data. This measurement yields the
value $\tau(B^+)=(1.68\pm0.07\pm0.02)\ {\rm ps}$, and a result of
comparable precision is obtained for the $B^0$ lifetime.

The third method for measuring $b$ hadron lifetimes makes use of
topological vertexing, in which the $B$ meson charge is determined
from the total charge of the tracks associated with the decay
vertex. Using this technique, SLD~\cite{SLD_lifetimes} has obtained 
6033 charged vertices and 3665 neutral vertices in 
a sample of only
150 thousand hadronic $Z$ decays. 
Their measurement errors are comparable to those of many of the
LEP and CDF results.

\begin{figure}[tbh]
\epsfxsize=3.5in
\hbox{\hfill
\epsffile{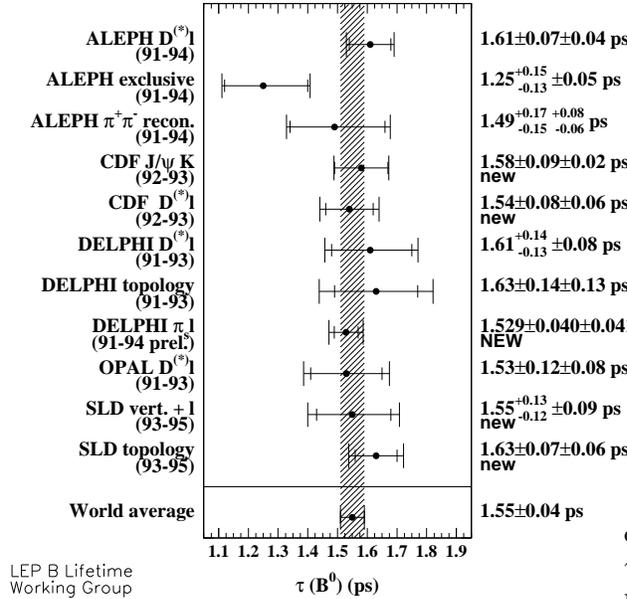}\hfill}
\vskip -2mm
\caption{Summary of $B^0$ lifetime 
measurements.\hspace*{\fill}}
\label{fig:b0_lifetimes}
\end{figure}

\begin{figure}[tbh]
\epsfxsize=3.5in
\hbox{\hfill
\epsffile{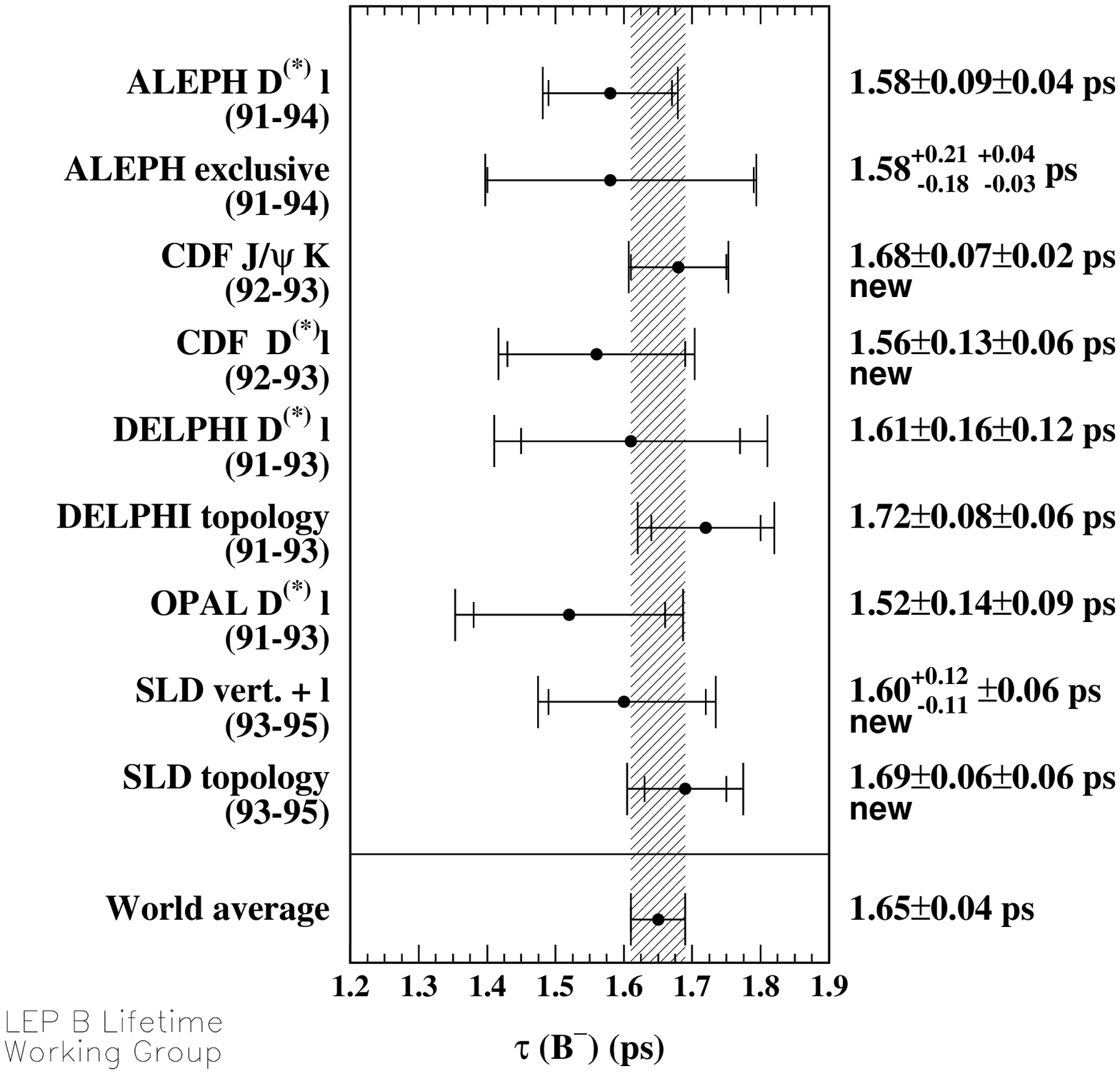}\hfill}
\vskip -2mm
\caption{Summary of $B^-$ lifetime 
measurements.\hspace*{\fill}}
\label{fig:bp_lifetimes}
\end{figure}

Figures~\ref{fig:b0_lifetimes} and \ref{fig:bp_lifetimes} list
the measured $B^0$ and $B^-$ lifetimes and give the world
averages computed by the LEP $B$ Lifetime 
Working Group.~\cite{LEP_Working} 
Their averages take into account the many sources of correlated
experimental error, such as assumed fragmentation models,
decay models, and branching fractions. 
To compare the $B^0$ and $B^-$
lifetimes, it is best to calculate the lifetime ratio for each experiment,
so that many systematics cancel, and then to compute
the average of the individual lifetime ratios. This average 
is 
\begin{eqnarray}
\tau(B^-)/\tau(B^0)=1.04\pm0.04, 
\end{eqnarray}
consistent with unity.
%
\begin{figure}[tbh]
\epsfxsize=3.5in
\hbox{\hfill
\epsffile{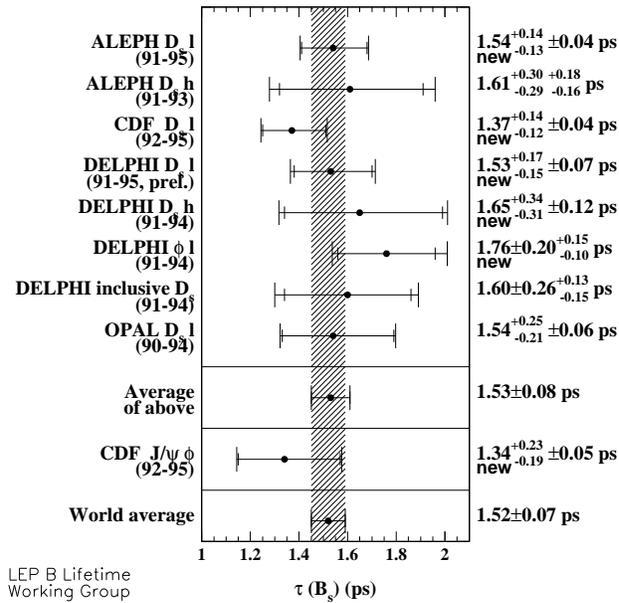}\hfill}
\vskip -2mm
\caption{Summary of $B_s$ 
lifetime measurements.\hspace*{\fill}}
\label{fig:bs_lifetimes}
\end{figure}

Figure~\ref{fig:bs_lifetimes} summarizes
measurements of the $B_s$ lifetime.
The uncertainties here are generally much larger
than those for $\bar B^0$ and $B^-$ lifetimes,
due to smaller event samples and larger
combinatorial backgrounds. Most of
the measurements use the semileptonic decay
$B_s\to D_s^+ X\ell^-\bar\nu$ and reconstruct
the $D_s^+$ decay in one or more modes, although some
use an inclusive $D_s^+$ signal or a $D_s^+$ with an 
associated hadron. The average lifetime from
measurements based on these modes
is $\tau(B_s)=1.53\pm0.08,$ consistent with the
$B^0$ and $B^-$ lifetimes.

Because the $B_s^0$ and $\bar B_s^0$ can
oscillate into one another, the
two mass eigenstates are expected to differ somewhat in
their masses and lifetimes. Theoretical estimates
indicate that the difference in decay widths may
be as large as $\Delta\Gamma/\Gamma=30\%$.~\cite{DeltaGamma_ref}
CDF~\cite{CDF_Bslife} has reconstructed 
$58\pm8$ events in the decay
$B_s\to J/\psi\phi$, which is expected 
to be predominantly CP even, in contrast to
the $D_s^+\ell^-X$ final state, which is expected
to be an equal mixture of CP even and CP odd. 
The preliminary result, 
$\tau(B_s)=1.34^{+0.23}_{-0.19}\pm0.05\ {\rm ps},$
is not yet sufficiently precise to address
this issue, but future measurements with more data should be
extremely interesting. More details on measurements of
$B^0,$ $B^-,$ and $B_s$ lifetimes may be found in the talk of 
Claire Shepherd-Themistocleous.~\cite{Claires_talk}
\begin{figure}[tbh]
\epsfxsize=3.5in
\hbox{\hfill
\epsffile{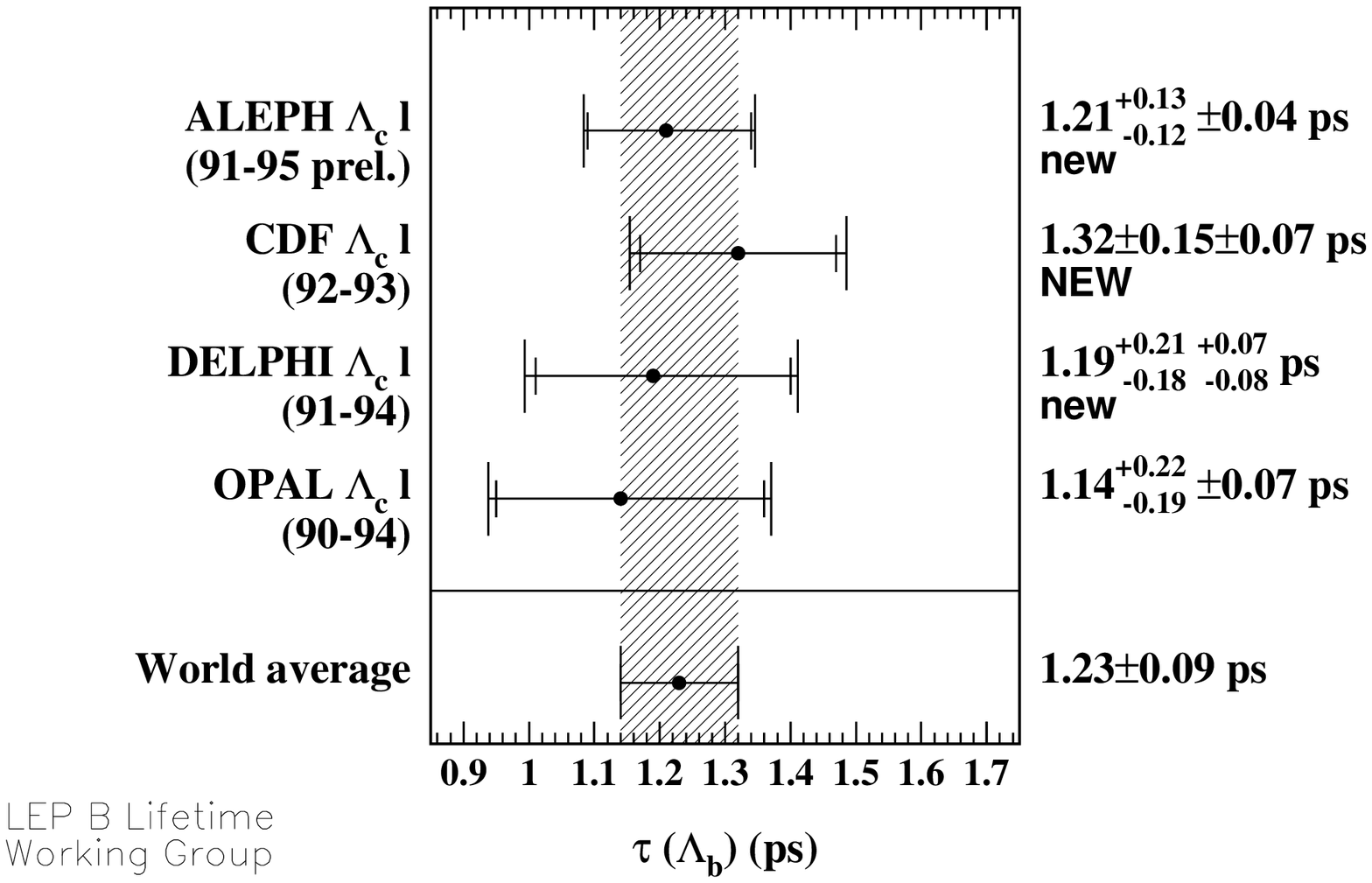}\hfill}
\vskip -2mm
\caption{Summary of $\Lambda_b$ lifetime
measurements.\hspace*{\fill}}
\label{fig:lambdab_lifetime}
\end{figure}

The measured values of $b$-baryon lifetimes 
are systematically lower than those for $B$ mesons. 
These analyses, which are described in more detail in the talk by 
Peter Ratoff~\cite{Ratoff_talk}, are performed using a
variety of methods that select different compositions of 
$b$ baryons, whose production fractions at the $Z$ or
at the Fermilab collider are not well known. 
(In fact, some papers use the symbol $\Lambda_b$ to
denote a generic $b$-baryon, but I use 
$\Lambda_b^0$ to denote the $I=0$ $bud$ ground state baryon.)
Measurements of the $\Lambda_b^0$ lifetime generally use 
$\Lambda_c^{\pm}\ell^{\mp}$ correlations to select events from the
decay $\Lambda_b^0\to\Lambda_c^+ X\ell^-\bar\nu$. 
By fully reconstructing the $\Lambda_c^+\to p K^-\pi^+$ 
decay and using kinematic
cuts, one can obtain a high purity
sample of $\Lambda_b^0\to\Lambda_c^+\ell^-\bar\nu$ decays. 
Alternatively, one can obtain larger event samples by using
the inclusive decay
$\Lambda_c^+\to\Lambda X$ and searching for $\Lambda\ell^-$ 
or even $p\ell^-$ combinations.
These methods, however, have a lower purity of $\Lambda_b^0$ baryons,
since, for example, there can be background from
$\Xi_b\to\Xi_c\ell^-\bar\nu$, $\Xi_c\to\Lambda X$. 

Figure~\ref{fig:lambdab_lifetime} shows the $\Lambda_b^0$
lifetime measurements obtained from $\Lambda_c^{\pm}\ell^{\mp}$ correlations
in which the $\Lambda_c^+$ is fully reconstructed. The world
average, $\tau_{\Lambda_B}=(1.23\pm0.09)$ ps, differs from the $B^0$
lifetime by about $3\sigma.$
By including measurements that use
$\Lambda\ell^-$ and $p\ell^-$ correlations, the LEP $B$ Lifetime
Working Group obtains the average $\tau_{b{\rm -baryon}}=(1.21\pm0.06)$ ps,
or $\tau_{b{\rm -baryon}}/\tau_{B^0}=0.78\pm0.04.$ This ratio is
difficult to accomodate in calculations based on the $1/m_Q$
expansion, which predict a value in the range 0.9 to 1.0.

OPAL~\cite{OPAL_baryon_sl} has measured the ratio $R_{\Lambda\ell}=
B(b{\rm -baryon}\to\Lambda\ell X)/B(b{\rm -baryon}\to\Lambda X)$,
which should be a good approximation 
to the average $b$ baryon semileptonic branching fraction.
They obtain $R_{\Lambda\ell}=(6.8\pm1.3\pm1.0)\%$,
which is consistent with the expectation based on the
shorter $b$ baryon lifetime and the assumption of equal
semileptonic widths of $B$ mesons and $b$ baryons. 
\section{Conclusions}

The combination of new measurements and new theoretical methods is
transforming the subject of heavy-flavor dynamics. For many 
such processes, we have much more than a qualitative understanding:
detailed, quantitative comparisons between theory and experiment have been
performed, and the results are quite encouraging. 
In particular, there has been substantial progress on decay constants
and semileptonic decay form factors. Quantities related to the hadronic rate,
such as the inclusive semileptonic branching fraction, $n_c$, and 
$b$-hadron lifetimes, have proved to be more difficult to understand
and have produced some intriguing puzzles. With the intensive effort
underway at several laboratories, there is every reason to expect that
these questions can be explored in great detail, and that many new
ones will arise as rare decay modes become more accessible.
\section*{Acknowledgements} Many people contributed 
to this talk by providing plots, answering questions, and
making suggestions on the presentation. I 
gratefully acknowledge the help of Thomas Browder, Andrzej Buras,
John Carr, Lawrence Gibbons, I.~Joseph Kroll, 
Guido Martinelli, Clara Matteuzzi,
Ritchie Patterson, Peter Ratoff, Claire Shepherd-Themistocleous, 
Su Dong, and the LEP $B$-lifetimes working group. This work was 
supported by the U.S.~Dept.~of Energy grant DE-FG03-91ER40618.

\section*{Questions}
\noindent{\it Yung su Tsai, SLAC:}

Will there ever be enough semileptonic decay events of $B$ and $D$,
so that we can measure the polarization of $\mu$ and $\tau$?
If polarization is different for charged conjugate decay modes, it
will indicate the non-standard model CP violation, i.e., the existence of
a new charged gauge boson.

\vskip 12pt
\noindent{\it J.~Richman:}

It is not possible to measure the $\mu$ polarization, since these
particles do not decay in the detector volume. In principle, one
could measure the $\tau$ polarization in the decay 
$\bar B\to D^*\tau^-\bar\nu,$ but it wouldn't be easy, and
it would require a very large data sample. While
this decay has been observed at LEP, it has not been seen at the
$\Upsilon(4S),$ since the presence of two neutrinos (including
the $\tau^-$ decay) makes its separation from background very difficult.

\vskip 12pt
\noindent{\it Bennie F.~L.~Ward, U.~of Tennessee:}

Could you explain why you say that the branching fraction $B(B\to D_sX)$
is experimentally different at CLEO and LEP, even though theoretically
a particle's branching fraction to a given final state cannot depend
on its velocity?

\vskip 12pt
\noindent{\it J.~Richman:}

This question is based on a misunderstanding. At LEP, the 
$b$-hadron mixture includes $B_s$ mesons and $b$ baryons in addition
to $B^+$ and $B^0$ mesons. Measurements of $D_s^+$ production
in this $b$-hadron mixture will therefore yield somewhat different
results than measurements at the $\Upsilon(4S)$, where the
$b$-hadron mixture includes only $B^0$ and $B^+$. 

\end{document}